\documentclass[12pt,preprint]{aastex}

\usepackage{amsmath}
\usepackage{amsfonts}
\usepackage{amssymb}
\usepackage{float}
\usepackage{graphicx}
\usepackage{color}
\usepackage{multirow}

\newcommand{\sgn}{\mathrm{sgn}}

\newcommand{\sK}{{\mathbb K}}

\newcommand{\bx}{\mathbf{x}}
\newcommand{\bvel}{\mathbf{v}}
\newcommand{\bp}{\mathbf{p}}
\newcommand{\bk}{\mathbf{k}}
\newcommand{\bE}{\mathbf{E}}
\newcommand{\bB}{\mathbf{B}}
\newcommand{\bj}{\mathbf{j}}
\newcommand{\be}{\mathbf{e}}

\newcommand\cm{{\rm\,cm}}

\title{Numerical Evaluation of the Relativistic Magnetized Plasma Susceptibility Tensor and Faraday Rotation Coefficients}

\author{Alex Pandya}
\affil{Department of Physics, Princeton University, Princeton, NJ, 08544, and 
Department of Physics, University of Illinois, 1110 West Green Street, Urbana, IL, 61801}

\author{Mani Chandra}
\affil{Research Division, Quazar Technologies, Sarvapriya Vihar, New Delhi, 110016}

\author{Abhishek Joshi}
\affil{Department of Physics, University of Illinois, 1110 West Green Street, Urbana, IL, 61801}

\author{Charles F. Gammie}
\affil{Department of Physics and Department of Astronomy, University of Illinois, 1110 West Green Street, Urbana, IL, 61801}

\begin{document}

\begin{abstract}

Polarized models of relativistically hot astrophysical plasmas require transport coefficients as input: synchrotron absorption and emission coefficients in each of the four Stokes parameters, as well as three Faraday rotation coefficients.  Approximations are known for all coefficients for a small set of electron distribution functions, such as the Maxwell-J\"uttner relativistic thermal distribution, and a general procedure has been obtained by Huang \& Shcherbakov for an isotropic distribution function.   Here we provide an alternative general procedure, with a full derivation, for calculating absorption and rotation coefficients for an arbitrary isotropic distribution function.  Our method involves the computation of the full plasma susceptibility tensor, which in addition to absorption and rotation coefficients may be used to determine plasma modes and the dispersion relation.  We implement the scheme in a publicly available library\footnotemark with a simple interface, thus allowing for easy incorporation into radiation transport codes. We also provide a comprehensive survey of the literature and comparison with earlier results.   

\end{abstract}

\footnotetext{\tt{https://github.com/afd-illinois/symphony}}

\section{Introduction}

The Event Horizon Telescope (EHT) is a millimeter wavelength Very Long Baseline interferometry collaboration that aims to resolve the event horizon of the low accretion rate black holes at the center of the Milky Way (Sgr A*) and M87 (\cite{doeleman2009}). EHT will produce resolved, polarized, time-dependent data for both sources.  This data will constrain models of the black hole accretion flow and any outflows or jets, the magnetic field geometry in the source, the state of the plasma, and possibly the black hole spacetime.  Still, interpreting the data will require models that accurately predict the resolved, polarized radiation field from a dynamical model for the accretion flow.  Our goal here is to narrow the uncertainties in the production of synthetic polarized maps of black hole accretion flows from underlying flow models.   

Black holes that are accreting at a sufficiently low rate, in the sense that the luminosity is small compared to the Eddington luminosity $L_{Edd} = 4 \pi G M c/\kappa_{es}$ ($M \equiv$ black hole mass, $\kappa_{es} \equiv$ electron scattering opacity), are believed to be surrounded by an optically thin, geometrically thick, magnetized disk (\cite{yuan2014}). Both M87 and Sgr A* are believed to be in this regime.  A geometrically thick disk must be relativistically hot close to the innermost stable circular orbit, since scale height $H$ and local radius $r$ are related through hydrostatic equilibrium by $(H/r)^2 = r \Theta_p c^2/(G M)$ ($\Theta_p \equiv k T_p/(m_p c^2)$ is the dimensionless proton temperature; $T_p \equiv$ proton temperature; $m_p \equiv$ proton mass). If the electrons are relativistic, the disk is collisionless if it is optically thin to Thomson scattering; hence $T_e = T_p$ ($T_e \equiv$ electron temperature) is not required, nor do protons and electrons need to follow a thermal distribution function.  Existing EHT observations resolve both Sgr A* and M87 and are consistent with dimensionless electron temperature $\Theta_e \equiv k T_e/(m_e c^2) \sim 10$ close to the innermost stable circular orbit (\cite{doeleman2008}).

Electrons in a magnetized plasma emit and absorb photons by the cyclo-synchrotron process. Synchrotron radiation is, in general, linearly and circularly polarized.  Recall that polarized radiation can be described by the Stokes vector $I_S = \{I, Q, U, V\}^T$, where $I$, which is positive definite, is total intensity, $Q$ and $U$ are signed and describe linear polarization with electric vector polarization angle (EVPA) at angle $\pi/4$ to each other,  and $V$ is signed and describes circular polarization.  A magnetized plasma can also induce generalized Faraday rotation, or Faraday conversion, that interconverts Stokes $Q, U,$ and $V$.  

Emission, absorption, and generalized Faraday rotation along a ray parameterized by a coordinate $s$ are governed by the polarized radiative transfer equation
\begin{equation}
\frac{d}{ds} I_{S}  = J_{S} - M_{ST} I_{T}.
\end{equation}
The vector $J_{S} = \{j_{I}, j_{Q}, j_{U},j_{V}\}^{T}$ contains the emission coefficients for each of the Stokes parameters.   The Mueller matrix is defined to be
\begin{equation}
M_{ST} = 
\begin{pmatrix}
\alpha_{I}    &\alpha_{Q}    &\alpha_{U}    &\alpha_{V}    \\
\alpha_{Q}    &\alpha_{I}    &\rho_{V}      &-\rho_{U}       \\
\alpha_{U}    &-\rho_{V}     &\alpha_{I}    &\rho_{Q}         \\
\alpha_{V}    &\rho_{U}      &-\rho_{Q}     &\alpha_{I}    
\end{pmatrix}.
\end{equation}
Here $\alpha_{S}$ are the absorption coefficients and $\rho_S$ are the generalized Faraday rotation coefficients (also called rotativities; $\rho_I$ does not exist).  Altogether there are 11 transfer coefficients: 4 emissivities, 4 absorptivities, and 3 rotation coefficients.  The covariant polarized transfer equation is described, with references to the relevant literature, in \cite{dexter2016} and \cite{moscibrodzka2017}.  These transfer coefficients may be related to components of the dielectric tensor (and the susceptibility tensor via equation \ref{eq:K_and_chi}), provided that the antihermitian part of the dielectric tensor is small compared to the hermitian part (see \cite{zheleznyakov1996} pg. 185-187).

A general procedure for calculating emissivities and absorptivities for a gyrotropic distribution function is provided in the publicly available code {\tt symphony}$^1$ \citep{pandya2016}, along with a comparison to other results in the literature.  Approximate formulae for all coefficients are provided in \cite{dexter2016}.   A general procedure for calculating rotativities for an isotropic distribution function was first provided by \cite{huang2011} via a mathematica script\footnote{\tt https://astroman.org/Faraday\_conversion/}.  In this paper we provide an alternative to the \cite{huang2011} approach to calculating rotativities and absorptivities with the aim of simplicity, transparency, and computational speed, so that our work may be immediately useful to those modeling radiative transfer.  Our results agree with \cite{huang2011}.   We also provide a survey of the literature, complete checkable derivations, and a publicly available C code with python interfaces.  These features are incorporated in the {\tt symphony} code.

The plan of the paper is as follows.  In \S 2 we define the susceptibility tensor and review the relations between its components and the components of the Mueller matrix.  In \S 3 we provide a general expression for the susceptibility tensor.  \S 4 describes a numerical scheme for evaluating the tensor, and \S 5 summarizes and compares to earlier work.  An Appendix (\S \ref{section:appendix}) provides a complete derivation of the results beginning with the linearized Vlasov equation.  

\section{Review and Definitions} \label{section:review_defns}

The components of the Mueller matrix are directly related to the classical linear response of the plasma to an imposed electromagnetic wave.  
We assume the wave has a time-varying electric field $\bE(t, \bx)$ that is turned on at $t = 0$.  We then have the transform
\begin{equation} \label{eq:E_transform}
E_j(\omega, \bk) = \int dx \, e^{-i \bk \cdot \bx } \int_{0}^{\infty} dt \, e^{i \omega t} \, E_{j}(t, \bx),
\end{equation}
which is a Fourier transform in $\bx$ and a Laplace transform in $t$.  The latter implies that $\omega$ is complex, and requires sufficiently large $\mathrm{Im}(\omega) > 0$ such that the transform converges as $t \to \infty$.  For real frequencies $\omega$, the standard approach involves making $\mathrm{Im}(\omega)$ infinitesimal and taking the limit $\mathrm{Im}(\omega) \to 0$ at the end of the calculation. 

We are most interested in the regime where $|\mathrm{Im}(\bk)| \ll |\mathrm{Re}(\bk)|$; if this condition is not met, the absorption length scale is on or near the order of the wavelength of the radiation, and the radiation will not be detectable after propagating through an astrophysical source kilometers or larger in size. Our treatment does not enforce this restriction on the wavevector, however, and our derivation is valid for complex $\bk$.

The plasma response can be expressed in terms of the plasma conductivity 3-tensor $\sigma_{ij}$ (units in Gaussian-cgs: $\sec^{-1}$), where
\begin{equation}
J_i = \sigma_{ij} E_j.
\end{equation}
Here $J_i$ is the induced current density.  Equivalently, the plasma response can be described by the dielectric 3-tensor $K_{ij}$ (dimensionless), where
\begin{equation}
D_i = \epsilon_0 K_{ij} E_j.
\end{equation}
Here $D_i \equiv$ induced displacement field, and $\epsilon_0 \equiv$ permittivity of free space ($= (4\pi)^{-1}$ in Gaussian-cgs).  The plasma response can also be described by the response 3-tensor $\alpha_{ij}$ (units in Gaussian-cgs: $\sec^{-1} \cm^{-1}$) in the temporal gauge\footnote{In which the scalar potential $\phi \rightarrow 0$, also known as the Hamiltonian or Weyl gauge.}, where
\begin{equation}
J_i = \alpha_{ij} A_j.
\end{equation}
Here $A_i$ is the imposed vector potential.  The relationship between these 3-tensors is 
\begin{align}
K_{ij} &= \delta_{ij} + \frac{i}{\omega \epsilon_0} \sigma_{ij} \label{eq:K_and_sigma}   \\
       &= \delta_{ij} + \frac{c}{\omega^2 \epsilon_0} \alpha_{ij} \label{eq:K_and_alpha} \\
       &= \delta_{ij} + \chi_{ij} \label{eq:K_and_chi},
\end{align}
(see, e.g., equation 6.17 from \cite{melrose1991}, except a missing factor of $c$ is inserted to correct equation \ref{eq:K_and_alpha}).   Here $\chi_{ij}$ is the dimensionless plasma susceptibility 3-tensor.   

A sufficient condition for the response of the plasma to be consistent with Maxwell's equations is
\begin{equation} \label{eq:k_eigenproblem}
\left(k_i k_j + \Big( \frac{\omega^2}{c^2} - k^2 \Big) \delta_{ij} + \frac{\omega^2}{c^2} \chi_{ij}(\omega,\bk) \right) E_j = 0.
\end{equation}
This is an eigenproblem with eigenvalues $\bk_A$ and eigenvectors $\bE_A$ for each plasma mode $A$.  At high frequency, where $|\chi_{ij}| \ll 1$,  the plasma supports two modes that, to first order in $|\chi_{ij}|$, depend only on the components of $\chi$ in the plane perpendicular to $\bk$.   

The relationship between $\chi_{ij}$ and the transfer coefficients is as follows.   If directions $1,2,3$ form a righthanded coordinate system and the wavevector points along the positive $3$ axis, then (see figure \ref{fig:dielectric_tensor_coord_diagram})
\begin{align}
\alpha_I &= \frac{2 \pi \omega \varepsilon_0}{ c } \, \mathrm{Im} \big( \chi^{\mathrm{ROT}}_{11} + \chi^{\mathrm{ROT}}_{22} \big) \label{eq:alpha_I}\\
\alpha_Q &= \frac{2 \pi \omega \varepsilon_0}{ c } \, \mathrm{Im} \big( \chi^{\mathrm{ROT}}_{11} - \chi^{\mathrm{ROT}}_{22} \big) \label{eq:alpha_Q}\\
\alpha_U &= \frac{2 \pi \omega \varepsilon_0}{ c } \, \mathrm{Im} \big( \chi^{\mathrm{ROT}}_{21} + \chi^{\mathrm{ROT}}_{12} \big) \label{eq:alpha_U}\\
\alpha_V &= \frac{2 \pi \omega \varepsilon_0}{ c } \, \mathrm{Re} \big( \chi^{\mathrm{ROT}}_{12} - \chi^{\mathrm{ROT}}_{21} \big) \label{eq:alpha_V}
\end{align}
and 
\begin{align}
\rho_Q   &= \frac{2 \pi \omega \varepsilon_0}{ c } \, \mathrm{Re} \big( \chi^{\mathrm{ROT}}_{22} - \chi^{\mathrm{ROT}}_{11} \big) \label{eq:rho_Q}\\
\rho_U   &= \frac{2 \pi \omega \varepsilon_0}{ c } \, \mathrm{Re} \big( \chi^{\mathrm{ROT}}_{21} + \chi^{\mathrm{ROT}}_{12} \big) \label{eq:rho_U} \\
\rho_V   &= \frac{2 \pi \omega \varepsilon_0}{ c } \, \mathrm{Im} \big( \chi^{\mathrm{ROT}}_{12} - \chi^{\mathrm{ROT}}_{21} \big), \label{eq:rho_V}
\end{align}
(\cite{sazonov1969, zheleznyakov1996, huang2011}).  The superscript ROT indicates that $\chi_{ij}$ is calculated in this wavevector-aligned coordinate system. Outside of this regime one must solve equation \ref{eq:k_eigenproblem} to compute transfer coefficients.

These relations are consistent with the relationship between the Stokes parameters and the polarization (or coherency) matrix given in \cite{zheleznyakov1996}, equation 1.54, and in, e.g. \cite{moscibrodzka2017} and \cite{huang2011}.   In particular, $Q > 0$ corresponds to linear polarization in the $1$ direction, $Q < 0$ to linear polarization in the $2$ direction, $U > 0$ to linear polarization along the $(\be_1 + \be_2)/\sqrt{2}$ axis, $U < 0$ to linear polarization along the $(\be_1 - \be_2)/\sqrt{2}$ axis, and $V > 0$ to right handed circular polarization according to the IEEE convention (see \cite{hamaker1996} for a discussion) in which the electric field vector rotates in a right-handed direction at fixed position if the thumb points in the direction of propagation (optical and infrared (OIR) astronomers typically use the opposite convention).    

The above relations are completely general.    We specialize to a magnetoactive plasma in which the magnetic field lies in the $1,3$ plane (see figure \ref{fig:dielectric_tensor_coord_diagram}).   Applying the Onsager relations (which result from the time-reversal invariance of the microscopic equations of motion) yields the result $\chi_{ij}(\textbf{B}) = \chi_{ji}(-\textbf{B})$ (\cite{stix1992}; \cite{melrose2008}).  This symmetry may be used to show that $\chi_{xy} = -\chi_{yx}$, $\chi_{zy} = -\chi_{yz}$, and $\chi_{xz} = \chi_{zx}$.  Following rotation into the Stokes basis, these results along with equations \ref{eq:alpha_U}, \ref{eq:rho_U}, and \ref{eq:chi_ROT_unsimplified} imply $\alpha_U = \rho_U = 0$.


\section{Susceptibility Tensor Calculation} \label{section:suscept_tensor_calculation}

\begin{figure}[h]
    \centering
    \includegraphics[scale=1.75]{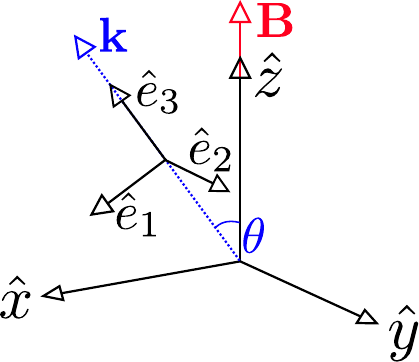}
    \caption{Diagram of the coordinate system used in the calculation.  The tensor $\chi_{ij}$ is computed in the $x,y,z$ basis, and the tensor $\chi^{\mathrm{ROT}}_{ij}$ is computed in the $1,2,3$ basis.  We define the $1$ direction by the component of $\hat{x}$ perpendicular to $\hat{e}_3$, and likewise the $2$ direction is defined by the component of $\hat{y}$ perpendicular to $\hat{e}_3$.  The wavevector $\bk$ lies in the $x$-$z$ plane.  Note: some authors (e.g. \cite{huang2011} and \cite{dexter2016}) keep the same definitions for the $1,2,3$ basis but instead choose the $\bk$ to be in the $y$-$z$ plane; this choice reverses the sign of their Stokes Q coefficients when compared to ours.}
    \label{fig:dielectric_tensor_coord_diagram}
\end{figure}

The full derivation of the susceptibility 3-tensor $\chi_{ij}$ for a magnetized plasma with isotropic particle distribution function is given in the Appendix (\S \ref{section:appendix}).  It is convenient to calculate $\chi_{ij}$ in a basis $x,y,z$ in which $\bB$ is aligned along the z axis, and obtain $\chi_{ij}^{\mathrm{ROT}}$ by rotation.   In particular (see Figure \ref{fig:dielectric_tensor_coord_diagram}) 
\begin{equation} \label{eq:rotation}
\chi^{\mathrm{ROT}} = R_y(\theta) \chi R_y^{T}(\theta)
\end{equation}
where 
\begin{equation}
R_y(\theta) \equiv 
\begin{pmatrix}
\cos (\theta)  & 0 & -\sin (\theta) \\
0            & 1 & 0           \\
\sin (\theta) & 0 & \cos (\theta)
\end{pmatrix}
\end{equation}
so that in the 1,2 plane
\begin{equation} \label{eq:chi_ROT_unsimplified}
\chi^{\mathrm{ROT}}_{ij} = 
\begin{pmatrix}
\cos^2 (\theta) \chi_{xx} - \cos (\theta) \sin (\theta) (\chi_{zx} + \chi_{xz}) + \sin^2 (\theta) \chi_{zz} & \cos (\theta) \chi_{xy} - \sin (\theta) \chi_{zy} \\
\cos (\theta) \chi_{yx} - \sin (\theta) \chi_{yz} & \chi_{yy}
\end{pmatrix},
\end{equation}
which simplifies using the Onsager symmetries noted above to 
\begin{equation}
\chi^{\mathrm{ROT}}_{ij} = 
\begin{pmatrix}
\cos^2 (\theta) \chi_{xx} - 2 \cos (\theta) \sin (\theta) \chi_{xz} + \sin^2 (\theta) \chi_{zz} & \cos (\theta) \chi_{xy} - \sin (\theta) \chi_{zy} \\
-(\cos (\theta) \chi_{xy} - \sin (\theta) \chi_{zy}) & \chi_{yy}
\end{pmatrix}.
\end{equation}

In order to compute $\chi_{ij}$, one must evaluate three momentum-space integrals over an integrand which includes the particle distribution function, as well as one integral over the unperturbed orbits of the particles.  In the results that follow we use a scaled version of the distribution function $\tilde{f} = m^3 c^3 f / n$, where the usual distribution function $f = dn/d^3 p$ with $p_i \equiv$ components of particle momenta.  

In brief, our approach involves the analytic evaluation of two of the three momentum-space integrals (one of which assumes that $f$ is isotropic), as well as the infinite Bessel function sum that arises as a result of one of these integrals.  The novel feature is that the remaining two-dimensional integral is numerically tractable and more physically transparent than the standard form of the susceptibility tensor.  We also provide a publicly available code to perform this numerical evaluation, which includes functions to compute the transfer coefficients $\alpha_S$ and $\rho_S$.

The final susceptibility 3-tensor for a single species with signed charge $q$, mass $m$, and number density $n$ has the form
\begin{equation} \label{eq:chi_ij_ours}
\chi_{ij} = \frac{2 \pi i \omega_p^2}{\omega \mathrm{Re}(\omega)} \int_{1}^{\infty} d\gamma \, (\gamma\beta)^3 \, \, \frac{d \tilde{f}}{d \gamma} \, \, \mathbb{K}_{ij}(\gamma, \omega/\omega_c, \bk),
\end{equation}
where the total susceptibility is obtained by summing the susceptibilities for each species.  The quantity $\omega_p^2 \equiv 4 \pi n q^2 / m$ is the species' plasma frequency, $\omega_c \equiv q B/(m c)$ is the cyclotron frequency, $\beta \equiv v/c$, $\gamma \equiv (1 - v^2/c^2)^{-1/2}$ is the Lorentz factor, $\theta \equiv \arccos(\hat{k} \cdot \hat{B})$ is the angle between the magnetic field and wavevector connecting the source to the observer, and $d\tilde{f}/d\gamma$ is the derivative of the (scaled) distribution function with respect to $\gamma$.  In the numerical evaluation of $\chi_{ij}$ this derivative is computed analytically to speed up evaluation.  These derivatives are:
\begin{align}
\frac{d \tilde{f}}{d\gamma} &= - \frac{\exp(-\gamma/\Theta_e)}{4 \pi \Theta_e^2 K_2(1/\Theta_e)} ~~~~~~~~~~~~~~~~~~~~~~~~~~~~(\text{Maxwell-J\"uttner}) \label{eq:MJ_deriv}\\
\frac{d \tilde{f}}{d\gamma} &= - \frac{(p - 1) (-1 + 2 \gamma^2 + p (\gamma^2 - 1))}{4 \pi (\gamma_{\mathrm{min}}^{-1 - p} - \gamma_{\mathrm{max}}^{-1 - p}) \beta (\gamma^2 - 1)} \gamma^{-3 - p} ~~~~~(\text{power-law}) \label{eq:PL_deriv}\\
\frac{d \tilde{f}}{d\gamma} &= - \frac{N_\kappa (1 + \kappa)}{\kappa w} \Big( 1 + \frac{\gamma - 1}{\kappa w} \Big)^{-2 - \kappa} ~~~~~~~~~~~~~~~~~~~(\text{kappa}). \label{eq:kappa_deriv}
\end{align}
Within equation \ref{eq:MJ_deriv}, $K_2$ is the second-order modified Bessel function of the second kind.  In equation \ref{eq:PL_deriv}, $p$ is the index of the power-law distribution function (the exponent on $\gamma$); $\gamma_{\mathrm{min}}, \gamma_{\mathrm{max}}$ are the lower and upper bounds (respectively) on $\gamma$ within which the distribution is nonzero. In equation \ref{eq:kappa_deriv}, $\kappa$ is the index parameter for the kappa distribution, $w$ is the width parameter of the kappa distribution, and $N_\kappa$ is the normalization constant, which is computed numerically.

Finally, the kernel $\mathbb{K}_{ij}$ is given by
\begin{equation} \label{eq:kernel_defn}
\mathbb{K}_{ij}(\gamma, \omega/\omega_c, \bk) = \int_{0}^{\infty} d\tau e^{i \gamma \frac{\omega}{\mathrm{Re}(\omega)} \tau} ~\Phi_{ij}(\tau, \gamma, \omega/\omega_c, \bk),
\end{equation}
where
\begin{multline}
\Phi_{ij}(\tau, \gamma, \omega/\omega_c, \bk) = \\
\begin{pmatrix}
-\frac{1}{2} [ \cos(\frac{\omega_c}{\mathrm{Re}(\omega)} \tau) \mathcal{I}_1(0) - \mathcal{I}_1(2)] & -\frac{1}{2} \sin(\frac{\omega_c}{\mathrm{Re}(\omega)} \tau) \mathcal{I}_1(0) & -\cos(\frac{\omega_c}{2 \mathrm{Re}(\omega)} \tau) \mathcal{I}_2(1) \\
- \Phi_{12} & -\frac{1}{2} [ \cos(\frac{\omega_c}{\mathrm{Re}(\omega)} \tau) \mathcal{I}_1(0) + \mathcal{I}_1(2) ] & \sin(\frac{\omega_c}{2 \mathrm{Re}(\omega)} \tau) \mathcal{I}_2(1) \\
\Phi_{13} & -\Phi_{23} & -\mathcal{I}_3(0)
\end{pmatrix}.
\end{multline}
The functions $\mathcal{I}(n, \tau, \gamma, \omega/\omega_c, \bk)$ (shown as $\mathcal{I}(n)$ above) are 
\begin{align}
\mathcal{I}_1(0) &= \frac{2 ((2 \alpha^2 + (\alpha^2 - 1) \delta^2 + \delta^4) \sin A - (2 \alpha^2 - \delta^2) A \cos A)}{A^5} \label{eq:I_1_of_0}\\
\mathcal{I}_1(2) &= - \frac{2 \delta^2 (3 A \cos A + (A^2 - 3) \sin A)}{A^5} \label{eq:I_1_of_2}\\
\mathcal{I}_2(1) &= \frac{ 2 i \alpha \delta \big( 3 A \cos A + (A^2 - 3) \sin A \big)}{A^5} \label{eq:I_2_of_1}\\
\mathcal{I}_3(0) &= \frac{6 \alpha^2 \cos A}{A^4} - \frac{2 \cos A}{A^2} + \frac{6 \delta^2 \sin A}{A^5} - \frac{4 \sin A}{A^3} + \frac{2 \alpha^2 \sin A}{A^3}, \label{eq:I_3_of_0}
\end{align}
where
\begin{align}
\alpha &= \frac{\gamma \beta c k \cos (\theta)}{\mathrm{Re}(\omega)} \tau \label{eq:alpha_defn}\\
\delta &= \frac{2 \gamma \beta c k \sin (\theta)}{\omega_c} \sin \Big( \frac{\omega_c}{2 \mathrm{Re}(\omega)} \tau \Big) \label{eq:delta_defn}\\
A &= \sqrt{\alpha^2 + \delta^2}. \label{eq:A_defn}
\end{align}
The quantity $k$ is the magnitude of the wavevector $\bk$.  Notice that our $\omega_c$ is a signed quantity, and is negative for electrons.  

When $|\mathrm{Im}(k)|$ is small, equation \ref{eq:chi_ij_ours} is well-behaved and convergent provided $\mathrm{Im}(\omega) > 0$; for real $\omega$, convergence is only lost when $\cos(\theta) = 0$, as the $\tau$ integrand becomes purely oscillatory.  We have not examined the convergence properties of equation \ref{eq:chi_ij_ours} for values of $k$ far from the real line, as these cases are outside of the astrophysically relevant regime $|\mathrm{Im}(k)| \ll |\mathrm{Re}(k)|$ (see \S \ref{section:review_defns} for a discussion).



\section{Numerical Algorithms}

Equation \ref{eq:chi_ij_ours} is free from singularities in both its real and imaginary parts, and no longer contains an infinite sum -- features which significantly complicate numerical evaluation of the standard version of the susceptibility tensor (see equation \ref{eq:textbook_chi} in \S \ref{section:appendix}).  However, the integrand in equation \ref{eq:chi_ij_ours} is oscillatory in both $\tau$ and $\gamma$.  Fortunately, if the integration over $\tau$ is performed first, the resultant integrand for $\gamma$ is smooth and rapidly convergent.  The rate-limiting step is the slowly-convergent $\tau$ integral, which is independent of all distribution function parameters, though it does depend on $\omega/\omega_c$ and $\bk$.

In the provided code we specialize to real $\omega$ and $k$, and throughout the remainder of this section we will use $\omega = \mathrm{Re}(\omega)$, $k = \mathrm{Re}(k)$.  In our algorithm we evaluate the two integrals serially, with the $\tau$ integration done first.  This process is slow, however, as the $\tau$ integral -- included in the kernel $\mathbb{K}_{ij}$ -- yields nonnegligible contributions at higher and higher $\tau$ as $\omega/\omega_c$ increases.  This behavior may be shown through analysis of the kernel's dependence on $\tau$.  The only terms in equation \ref{eq:kernel_defn} which decay in $\tau$ (for real $\omega, k$) do so because of inverse powers of $A$.  Pulling $\alpha \equiv \gamma \beta c k \cos (\theta) \tau / \omega$ out of the root in equation \ref{eq:A_defn} and writing $A$ out explicitly
\begin{equation*}
A = \frac{\gamma \beta c k \cos(\theta) \tau}{\omega} \sqrt{1 + 4 \tan^2(\theta) \frac{\omega^2}{\omega_c^2} \frac{\sin^2 ( \frac{\omega_c}{2 \omega} \tau )}{\tau^2}},
\end{equation*}
implying that the kernel only decays like $A \propto \tau$ when the second term in the root is small, namely for $\tau \gg 2 \tan (\theta) \, \omega / \omega_c $.  Thus in the large-$\tau$ limit: $\mathcal{I}_1(2) \propto 1/\tau^3$; $\mathcal{I}_1(0), \mathcal{I}_2(1) \propto 1/\tau^2$; $\mathcal{I}_3(0) \propto \sin(\alpha)/\alpha \sim \sin(\tau)/\tau$.  Despite the slow decay in $\tau$, all of these integrals are convergent as long as $\cos(\theta) \neq 0$; otherwise we must make use of the fact that $\mathrm{Im}(\omega) > 0$ (see the final paragraph of \S \ref{section:suscept_tensor_calculation}).

Since $\mathbb{K}_{ij}$ is smooth in $\gamma$ and independent of all external parameters except $\omega/\omega_c$ and $\bk$, it is possible to precompute and tabulate the kernel, which may be used to produce a fast spline fit to the $\gamma$ integrand.  This spline fit may then be integrated over $\gamma$ to yield a nearly instantaneous evaluation of the tensor for any isotropic distribution function.  In the module added to {\tt symphony}, we have provided spline fits to $\mathbb{K}_{ij}$ valid for the range $1 \leq \gamma, \omega/\omega_c \leq 1000$.

We provide functions to calculate both absorption and rotation coefficients from the susceptibility tensor.  We also provide the full $\tau$-$\gamma$ integrator so that the reader can access $\gamma, \omega/\omega_c$ values outside our precomputed intervals. 

\begin{figure}[h]
	\centering
	\includegraphics[width=\textwidth]{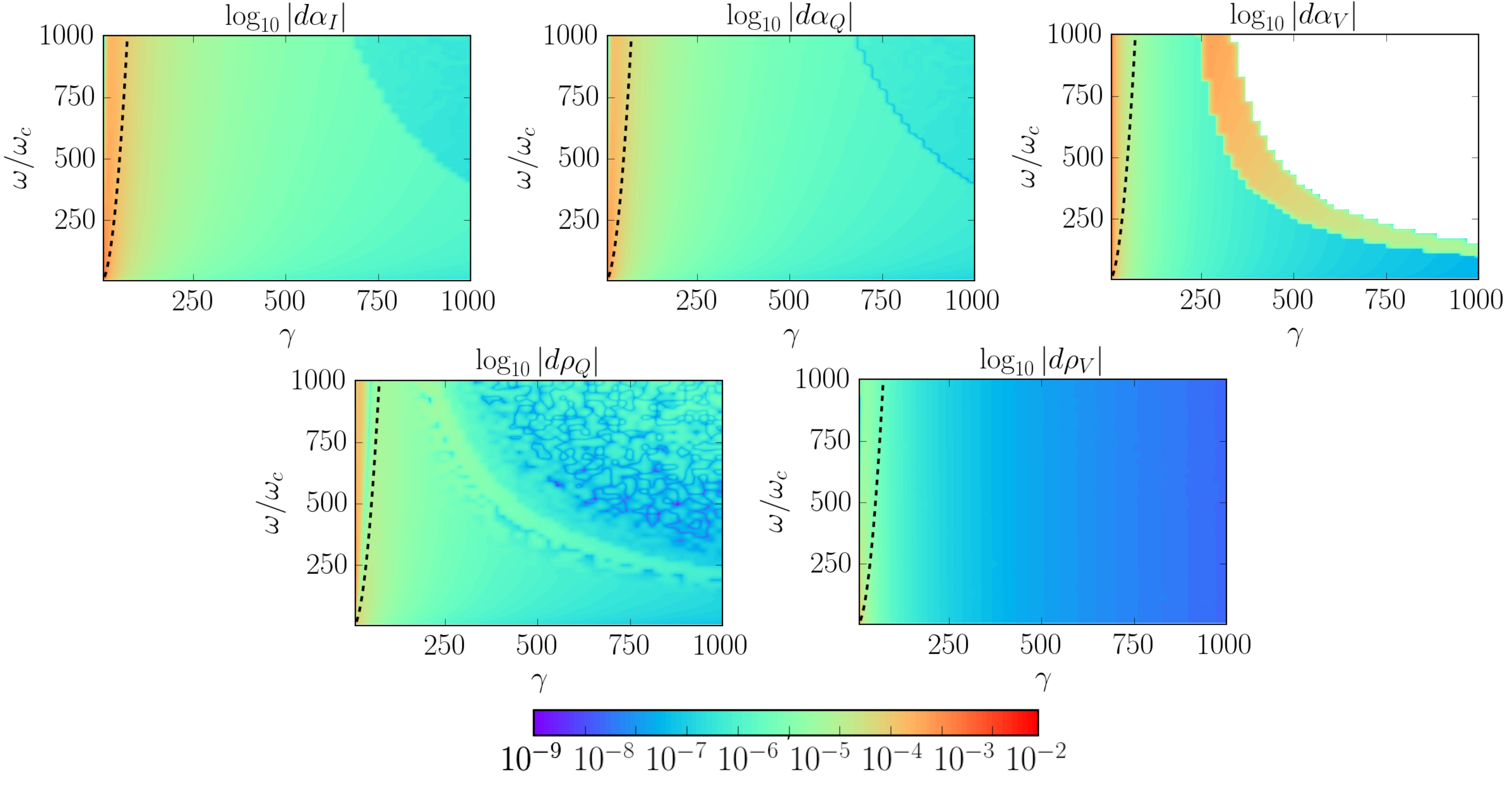}
	\caption{The kernel $\mathbb{K}_{ij}$, which is transformed into the Stokes basis by equations \ref{eq:alpha_I}-\ref{eq:rho_V} to yield the transfer coefficients prior to $\gamma$ integration.  We call these unintegrated coefficients $d\alpha_{I,Q,V}(\gamma,\omega/\omega_c,\theta)$ and $d\rho_{Q,V}(\gamma,\omega/\omega_c,\theta)$.  The coefficients $d\alpha_U, d\rho_U$ are equal to zero due to the Onsager relations, and are not shown.  White in the upper-right panel denotes a region where the coefficient $d\alpha_V$ is zero to within machine precision.  Cancellation between multiple oscillatory factors in the integrand produces the pattern in the upper-right corner of the $d\rho_Q$ plot.}
	\label{fig:d_alpha_and_d_rho}
\end{figure}

Figure  \ref{fig:d_alpha_and_d_rho} shows the $\sK_{ij}$ rotated and then transformed into the Stokes basis according to equations \ref{eq:alpha_I}-\ref{eq:rho_V}.   We call the resulting five coefficients $d\alpha_{I,Q,V}(\gamma,\omega/\omega_c,\theta)$ and $d\rho_{Q,V}(\gamma,\omega/\omega_c,\theta)$, as they comprise most of the integrand for these transfer coefficients prior to integration over $\gamma$.  Note that these coefficients depend on the observer angle $\theta$ rather than the full wavevector $\bk$; throughout the code we use the astrophysically relevant assumption $\omega \approx c k$ to eliminate the magnitude of the wavevector.  The figure shows the extent to which electrons contribute to the absorptivity and rotativity at a given $\gamma$ and $\omega/\omega_c$ for $\theta = \pi/3$, and the dashed line shows $\omega/\omega_c = (2/9) \gamma^2 \sin (\theta)$, the critical value of the  frequency where the interaction of the electron with the radiation field is expected to peak.

\section{Tests and Comparison to Earlier Work}

 We have performed a number of tests and comparisons to earlier work, not all of which we will describe in detail here, including: comparison of $\chi_{ij}$ for a thermal distribution in the nonrelativistic (NR) limit with the well-known warm and cold plasma $\chi_{ij}$; numerical comparison of $\chi_{ij}$ for a thermal distribution in the relativistic limit to the \cite{trubnikov1958} formulation of the same tensor; numerical comparison of absorption coefficients calculated from combinations of components of $\chi_{ij}$ for thermal, power-law, and kappa distributions with those computed using the alternative algorithm in {\tt symphony} (\cite{pandya2016}); and comparison of the Faraday rotation coefficients for the thermal distribution with the fitting formulae derived by \cite{huang2011} and extended by \cite{dexter2016}.  

To persuade the reader that our formulation is indeed correct, we first show explicitly that our formulation for the thermal distribution function is equivalent in the nonrelativistic limit to the well-known warm plasma susceptibility tensor, and then show numerical comparisons of the absorption coefficients with those from {\tt symphony}.

\subsection{Thermal Susceptibility Tensor NR Limit}
The nonrelativistic susceptibility tensor can be derived starting from either equation \ref{eq:chi_before_cosxi_integration} or the final susceptibility tensor (equation \ref{eq:chi_ij_ours}), each with the Maxwell-J\"uttner distribution function for $f$ (see equation \ref{eq:MJ_deriv}).  Focusing on the latter approach, one begins by taking the nonrelativistic limit $\beta \ll 1$, where the Maxwell-J\"uttner distribution function becomes the familiar Maxwell-Boltzmann distribution.  Converting the measure from from $\beta$ to the velocity $v$, we find that the $v$ integral is standard and is of Gaussian type with an additional factor of $v^n$ for some nonnegative integer $n$.  Evaluating this velocity space integral analytically yields
\begin{multline}\label{eq:warm_plasma_pandya}
\chi_{ij}^{MB} = - \frac{i \omega_p^2}{\omega \mathrm{Re}(\omega)} \int_{0}^{\infty} d \tau e^{i \frac{\omega}{\mathrm{Re}(\omega)} \tau} e^{- 2 \lambda \sin^2(\frac{\omega_c \tau}{2 \mathrm{Re}(\omega)})} e^{-\frac{w_T^2 k_z^2 \tau^2}{4 \mathrm{Re}(\omega)}^2} \\
\begin{pmatrix}
C (-1 + 2 \lambda (S_2)^2) + 2 \lambda (S_2)^2 & S (2 \lambda (S_2)^2 - 1) & - \frac{k_\perp k_z w_T^2 \tau}{\mathrm{Re}(\omega) \omega_c} C_2 S_2 \\
- S (2 \lambda (S_2)^2 - 1) & C (-1 + 2 \lambda (S_2)^2 ) - 2 \lambda (S_2)^2 & - \frac{ k_\perp k_z w_T^2 \tau}{\mathrm{Re}(\omega) \omega_c} (S_2)^2 \\
- \frac{k_\perp k_z w_T^2 \tau}{\mathrm{Re}(\omega) \omega_c} C_2 S_2 & \frac{ k_\perp k_z w_T^2 \tau}{\mathrm{Re}(\omega) \omega_c} (S_2)^2 & 1 - \frac{w_T^2 k_z^2 \tau^2}{2 \mathrm{Re}(\omega)^2}
\end{pmatrix},
\end{multline}
where $C \equiv \cos(\frac{\omega_c \tau}{\mathrm{Re}(\omega)})$, $C_2 \equiv \cos(\frac{\omega_c \tau}{2 \mathrm{Re}(\omega)})$, $S \equiv \sin(\frac{\omega_c \tau}{\mathrm{Re}(\omega)})$,  and $S_2 \equiv \sin(\frac{\omega_c \tau}{2 \mathrm{Re}(\omega)})$ are introduced to save space.  The quantity $\lambda \equiv \frac{k_\perp^2 w_T^2}{2 \omega_c^2}$, where $w_T = \sqrt{2 k_B T / m}$ is the (nonrelativistic) thermal speed, $k_\perp = \textbf{k} \cdot \hat{s} = |\textbf{k}| \sin (\theta)$ (where $\hat{s} \equiv (\hat{x} + \hat{y})/\sqrt{2}$ is the cylindrical radial coordinate) is the magnitude of the component of the wavevector perpendicular to the magnetic field, and $k_z = \textbf{k} \cdot \hat{z} = |\textbf{k}| \cos (\theta)$ is the magnitude of the component parallel to the field.  For $k_z = 0$ and $k$ real, one must use the fact that $\mathrm{Im}(\omega) > 0$ in order for the integral to converge (see \S \ref{section:review_defns} for further discussion).

Equation (\ref{eq:warm_plasma_pandya}) is superficially different from the warm plasma susceptibility tensor as given by Swanson\footnote{Here the tensor is corrected by a factor of $\sgn(q)$ in the components $\chi_{13}^{MB, ~Swanson} = \chi_{31}^{MB, ~Swanson}$, which is erroneously dropped in his derivation.  The missing sign is absorbed into our factors of $\omega_c$, which is signed here but unsigned in Swanson's work.} (2003; 2008):
\begin{multline}\label{eq:warm_plasma_swanson}
\chi_{ij}^{MB, ~Swanson} = \frac{\omega_p^2 e^{-\lambda}}{\omega k_z} \\
\begin{pmatrix}
\frac{1}{w_T} \sum\limits_{n=0}^{\infty} \frac{n^2 I_n}{\lambda} Z(\zeta_n) & \frac{i \, \sgn(q)}{w_T} \sum\limits_{n=0}^{\infty} n (I_n - I_n') Z(\zeta_n) & \frac{k_\perp}{2 \omega_c} \sum\limits_{n=0}^{\infty} \frac{n I_n}{\lambda} Z'(\zeta_n) \\
-\frac{i \, \sgn(q)}{w_T} \sum\limits_{n=0}^{\infty} n (I_n - I_n') Z(\zeta_n) & \sum\limits_{n=0}^{\infty} \frac{n^2 I_n}{w_T \lambda} Z(\zeta_n) + \frac{2 \lambda}{w_T} \sum\limits_{n=0}^{\infty} (I_n - I_n') Z(\zeta_n) & -\frac{i k_\perp}{2 \omega_c} \sum\limits_{n=0}^{\infty} (I_n - I_n') Z'(\zeta_n)\\
\frac{k_\perp}{2 \omega_c} \sum\limits_{n=0}^{\infty} \frac{n I_n}{\lambda} Z'(\zeta_n) & \frac{i k_\perp}{2 \omega_c} \sum\limits_{n=0}^{\infty} (I_n - I_n') Z'(\zeta_n) & - \frac{1}{w_T} \sum\limits_{n=0}^{\infty} I_n \zeta_n Z'(\zeta_n) \\
\end{pmatrix},
\end{multline}
where the function $I_n(\lambda)$ is the modified Bessel function of the first kind with argument $\lambda$ (suppressed to save space); $I_n'(\lambda) \equiv \partial I_n / \partial \lambda$ is its derivative; $\mathrm{sgn}(q)$ is the sign of the charge for the species in question ($= -1$ for electrons).  The function $Z(\zeta_n)$ is the plasma dispersion function, defined to be
\begin{equation} \label{eq:Z}
Z(\zeta) \equiv \frac{1}{\sqrt{\pi}} \int_{-\infty}^{\infty} \frac{e^{-\xi^2}}{\xi - \zeta} d\xi ~~~~~\mathrm{Im}(\zeta) > 0,
\end{equation}
with argument $\zeta_n = (\omega + n |\omega_c|)/(k_z w_T)$, and $Z'(\zeta) = -2 [1 + \zeta Z(\zeta)]$ is its derivative.  For $\mathrm{Im}(\zeta) \leq 0$, $Z$ is taken to be the analytic continuation of equation \ref{eq:Z}.  Note that the integral in the plasma dispersion function contains a simple pole at $\xi = \zeta$; applying the Sokhotski-Plemelj theorem allows one to rewrite the integral in terms of a purely real Cauchy principal value integral, plus a constant imaginary part.

Though equations \ref{eq:warm_plasma_pandya} and \ref{eq:warm_plasma_swanson} appear different, they are both derived starting with the same equation (see equation \ref{eq:stix_45} in \S \ref{section:appendix}); the difference comes in the next step, where the standard approach evaluates the $\tau$ integral resulting in a resonant denominator, which becomes the plasma dispersion function $Z(\zeta_n)$ above (see equation \ref{eq:resonant_denom_integral}).  We avoid this step and instead analytically evaluate the infinite sum, after which it is possible to analytically evaluate the two remaining momentum-space integrals, leaving only a single integral over $\tau$.

At this stage is it still possible to equate the two tensors analytically.  We do so for one component ($\chi_{31} = \chi_{13}$), and leave the remaining components as an exercise for the reader.  All of the techniques required to analytically equate the remaining components are shown in the derivation below.

\subsubsection{Analytic Comparison of $\chi_{13}^{MB}$}
Beginning with the component $\chi_{13}^{MB}$ of equation \ref{eq:warm_plasma_pandya}, we note that the term $\exp(-2\lambda \sin^2 (\frac{\omega_c \tau}{2 \mathrm{Re}(\omega)}))$ may be rewritten using the sine power-reducing identity $\sin^2(x) = \frac{1 - \cos(2x)}{2}$.  Applying the identity and simplifying, we arrive at
\begin{equation*}
e^{- 2 \lambda \sin^2(\frac{\omega_c \tau}{2 \mathrm{Re}(\omega)})} = e^{- \lambda} e^{i (-i \lambda) \cos(\frac{\omega_c \tau}{\mathrm{Re}(\omega)})}.
\end{equation*}
We may now apply the Jacobi-Anger identity and then a well-known Bessel function identity
\begin{align}
e^{i z \cos \theta} &= \sum_{n = -\infty}^{\infty} i^n J_n(z) e^{i n \theta} \\
J_n(-x) &= (-1)^n J_n(x) ~~~\mathrm{for~}n\in \mathbb{Z},
\end{align}
which yields
\begin{equation*}
e^{- 2 \lambda \sin^2(\frac{\omega_c \tau}{2 \mathrm{Re}(\omega)})} =  e^{- \lambda} \sum_{n = -\infty}^{\infty} i^{-n} J_n(i \lambda) e^{\frac{i n \omega_c \tau}{\mathrm{Re}(\omega)}}.
\end{equation*}
In this equation we may identify the definition of the modified Bessel function of the first kind
\begin{equation}
I_n(x) = i^{-n} J_n(i x)
\end{equation}
and finally arrive at
\begin{equation*}
e^{- 2 \lambda \sin^2(\frac{\omega_c \tau}{2 \mathrm{Re}(\omega)})} = e^{- \lambda} \sum_{n = -\infty}^{\infty} I_n(\lambda) e^{\frac{i n \omega_c \tau}{\mathrm{Re}(\omega)}}.
\end{equation*}
Substituting this result into the $\chi_{13}$ component of equation \ref{eq:warm_plasma_pandya}, we find
\begin{multline}
\chi^{MB}_{13} = - \frac{i \omega_p^2 k_\perp k_z w_T^2}{\omega \mathrm{Re}(\omega)^2 \omega_c} e^{- \lambda} \sum_{n = -\infty}^{\infty} I_n(\lambda) \\
\times \int_{0}^{\infty} d \tau~ \tau e^{i \frac{\omega}{\mathrm{Re}(\omega)} \tau} e^{\frac{i n \omega_c \tau}{\mathrm{Re}(\omega)}} e^{-\frac{w_T^2 k_z^2 \tau^2}{4 \mathrm{Re}(\omega)}^2} \cos \Big( \frac{\omega_c \tau}{2 \mathrm{Re}(\omega)} \Big) \sin \Big( \frac{\omega_c \tau}{2 \mathrm{Re}(\omega)} \Big). 
\end{multline}

We now use Feynman's trick\footnote{This technique is more formally known as Leibniz's rule for differentiation under the integral.} to remove the factor of $\tau$ from the integrand, making the integral one of Gaussian type:
\begin{multline}
\chi^{MB}_{13} = - \frac{i \omega_p^2 k_\perp k_z w_T^2}{\omega \mathrm{Re}(\omega)^2 \omega_c} e^{- \lambda} \sum_{n = -\infty}^{\infty} I_n(\lambda) \Big[\frac{\mathrm{Re}(\omega)}{i \omega_c} \frac{\partial}{\partial n} \Big] \\
\times \int_{0}^{\infty} d \tau~ e^{i \frac{\omega}{\mathrm{Re}(\omega)} \tau} e^{\frac{i n \omega_c \tau}{\mathrm{Re}(\omega)}} e^{-\frac{w_T^2 k_z^2 \tau^2}{4 \mathrm{Re}(\omega)^2}} \cos \Big( \frac{\omega_c \tau}{2 \mathrm{Re}(\omega)} \Big) \sin \Big( \frac{\omega_c \tau}{2 \mathrm{Re}(\omega)} \Big),
\end{multline}
which may be evaluated by hand after applying Euler's formula to write the sine and cosine as complex exponentials, or using a symbolic integration software.  Evaluating the integral and then the derivative with respect to $n$, we find
\begin{multline} \label{eq:chi_13_integrated}
\chi^{MB}_{13} = \frac{i}{2} \frac{\omega_p^2 k_\perp k_z w_T^2}{\omega \omega_c} e^{- \lambda} \sum_{n = -\infty}^{\infty} I_n(\lambda)  \\
\sqrt{\pi} \Bigg[ \frac{ ((n-1) \omega_c+\omega) e^{-\frac{((n-1) \omega_c+\omega)^2}{k_z^2 w_T^2}}}{k_z^3 w_T^3} \left(1 + i \, \text{erfi}\left(\frac{(n-1) \omega_c+\omega}{k_z w_T}\right)\right) \\
-\frac{((n+1) \omega_c+\omega) e^{-\frac{((n+1) \omega_c+\omega)^2}{k_z^2 w_T^2}}}{k_z^3 w_T^3} \left(1 + i \, \text{erfi}\left(\frac{(n+1) \omega_c+\omega}{k_z w_T}\right)\right) \Bigg],
\end{multline}
where $\mathrm{erfi}(x)$ is the imaginary error function of argument $x$, and all factors of $\mathrm{Re}(\omega)$ have canceled.  \cite{swanson2008} equations A.14-A.15 relate the imaginary error function to the plasma dispersion function via an intermediary function called $w(x)$:
\begin{align}
w(x) &= e^{-x^2} (1 + i \, \mathrm{erfi}(x)) \\
Z(x) &= i \sqrt{\pi} w(x),
\end{align}
which we can immediately identify in equation \ref{eq:chi_13_integrated} and then replace with the plasma dispersion function to arrive at
\begin{equation}
\chi^{MB}_{13} = \frac{1}{2} \frac{\omega_p^2 k_\perp k_z w_T^2}{\omega \omega_c} e^{- \lambda} \sum_{n = -\infty}^{\infty} I_n(\lambda) \frac{1}{k_z^2 w_T^2}\Big[ \zeta_{n-1} Z\left( \zeta_{n-1} \right) - \zeta_{n+1} Z\left( \zeta_{n+1} \right) \Big],
\end{equation}
where we have also identified $\zeta_n = (\omega + n |\omega_c|)/(k_z w_T)$.  Distributing the sum into the two terms, we may shift the sum indices on the first term such that $n \to n-1$, and on the second term $n \to n+1$; pulling out common terms yields
\begin{equation}
\chi^{MB}_{13} = \frac{1}{2} \frac{\omega_p^2 k_\perp k_z w_T^2}{\omega \omega_c} e^{- \lambda} \frac{1}{k_z^2 w_T^2} \sum_{n = -\infty}^{\infty} \zeta_{n} Z\left( \zeta_{n} \right) \Big[  I_{n+1}(\lambda) - I_{n-1}(\lambda) \Big].
\end{equation}
We now make use of another Bessel function identity
\begin{equation}
I_{n+1}(x) - I_{n-1}(x) = -\frac{2 n}{x} I_n(x)
\end{equation}
to write
\begin{equation}
\chi^{MB}_{13} = \frac{1}{2}\frac{\omega_p^2 k_\perp k_z w_T^2}{\omega \omega_c} e^{- \lambda} \frac{1}{k_z^2 w_T^2} \sum_{n = -\infty}^{\infty} \zeta_{n} Z\left( \zeta_{n} \right) \Big[ \frac{-2 n}{\lambda} I_n(\lambda) \Big],
\end{equation}
and another
\begin{equation}
\sum_{n = -\infty}^{\infty} n I_n = 0,
\end{equation}
to add on a term equal to zero
\begin{equation}
\begin{aligned}
\chi^{MB}_{13} &= \frac{1}{2} \frac{\omega_p^2 k_\perp k_z w_T^2}{\omega \omega_c} e^{- \lambda} \frac{1}{k_z^2 w_T^2} \sum_{n = -\infty}^{\infty} \zeta_{n} Z\left( \zeta_{n} \right) \Big[ \frac{-2 n}{\lambda} I_n(\lambda) \Big] + \frac{1}{2} \frac{\omega_p^2}{\omega} e^{- \lambda} \frac{1}{k_z^2 w_T^2} \sum_{n = -\infty}^{\infty} \Big[ \frac{-2 n}{\lambda} I_n(\lambda) \Big] \\
&= \frac{1}{2} \frac{\omega_p^2 k_\perp k_z w_T^2}{\omega \omega_c} e^{- \lambda} \frac{1}{k_z^2 w_T^2} \sum_{n = -\infty}^{\infty} (-2) [\zeta_{n} Z\left( \zeta_{n} \right) + 1] \frac{n I_n(\lambda)}{\lambda};
\end{aligned}
\end{equation}
we can now use the identity
\begin{equation}
Z'(\zeta) = -2 [1 + \zeta Z(\zeta)]
\end{equation}
to arrive at
\begin{equation}
\chi^{MB}_{13} = \frac{1}{2} \frac{\omega_p^2 k_\perp}{\omega \omega_c k_z} e^{- \lambda} \sum_{n = -\infty}^{\infty} \frac{n I_n(\lambda)}{\lambda} Z'(\zeta_n),
\end{equation}
which is Swanson's form of the susceptibility tensor component ($\chi_{13}^{MB, ~Swanson}$) for the nonrelativistic Maxwellian distribution (equation \ref{eq:warm_plasma_swanson}).

\subsection{Numerical Comparison} \label{section:numerical_comparison}
\begin{figure}[h]
	\centering
	\includegraphics[height=0.95\textheight]{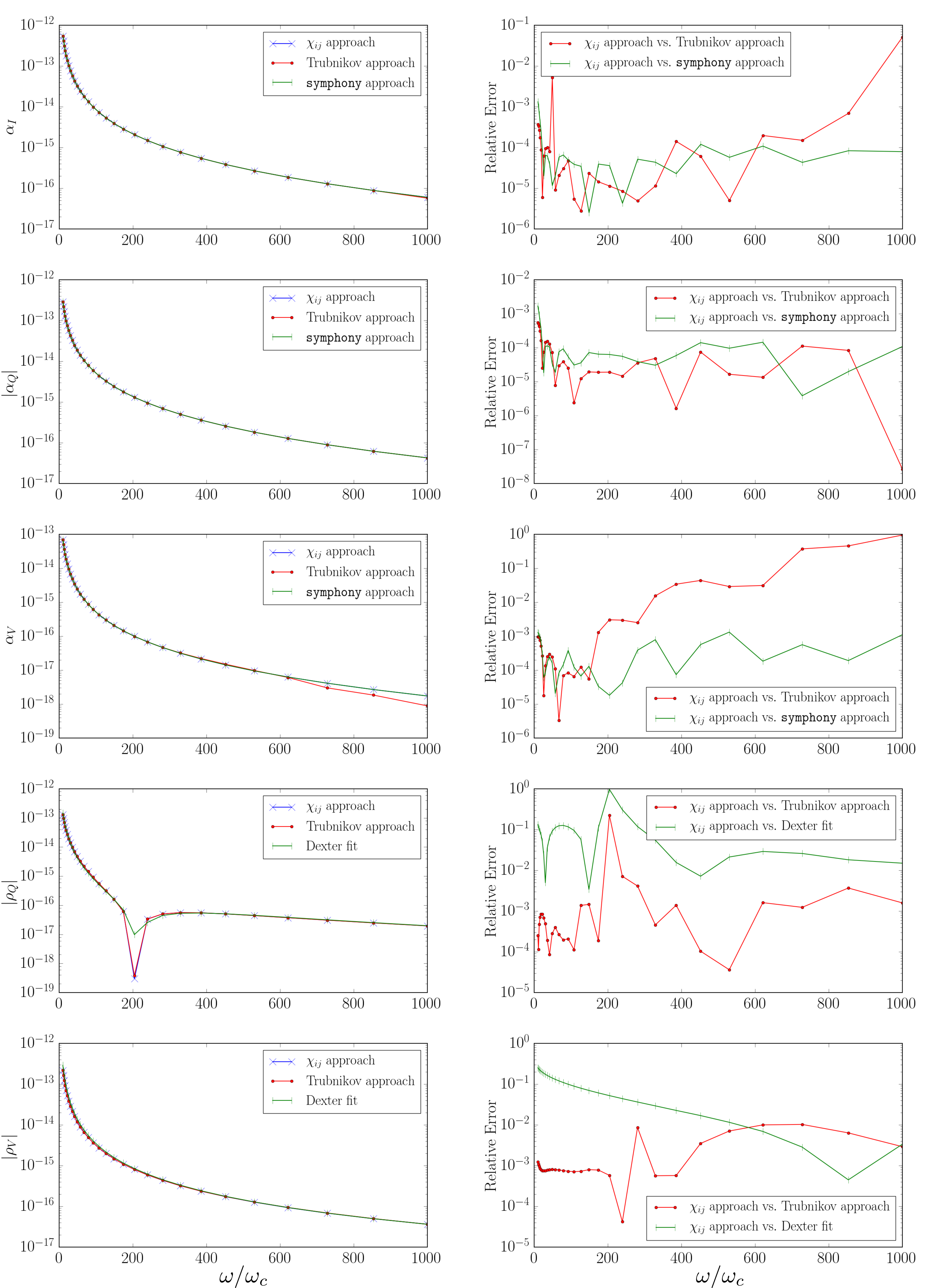}
	\caption{Figure showing the transfer coefficients $\alpha_S$ and $\rho_S$ (with $S \in \{I, Q, V\}$) for the Maxwell-J\"uttner distribution, along with alternative approaches to computing these quantities from the literature.  Relative error plots are shown on the right. See text \S \ref{section:numerical_comparison} for more discussion.}
	\label{fig:MJ_coeffs}
\end{figure}

To test our formulation of the susceptibility tensor in the relativistic limit (equation \ref{eq:chi_ij_ours}), we compute the transfer coefficients $\alpha_S$ and $\rho_S$ (with $S \in \{I, Q, V\}$; $\alpha_U = \rho_U = 0$ with our choice of coordinates) and compare the results from our approach to existing methods in the literature.  For the absorption coefficients $\alpha_S$, we compare our code's output (labeled ``$\chi_{ij} ~\mathrm{approach}$" in figure \ref{fig:MJ_coeffs}) to the result of numerically integrating the relativistic thermal susceptibility tensor derived by \cite{trubnikov1958} and to $\alpha_S$ as computed by the alternative algorithm in {\tt symphony}.  For the Faraday rotation coefficients $\rho_S$, we compare our approach again to that of Trubnikov, and to the fitting formulae supplied by \cite{dexter2016} and \cite{huang2011}.  For all plots in figure \ref{fig:MJ_coeffs} we choose fiducial parameters $\theta = \pi/3$, $\Theta_e = 10$, and for the error plots (shown on the right-hand side of each corresponding plot) we compute the error as follows:
\begin{equation}
\mathrm{Relative~Error} = \frac{|\mathrm{our~approach} - \mathrm{standard~approach}|}{\mathrm{standard~approach}}.
\end{equation}

We find agreement to within 1\% for most coefficients across the sampled range in $\omega/\omega_c$.  Large errors when compared to Trubnikov's tensor arise due to difficulty numerically integrating Trubnikov's tensor at high frequency, as it oscillates more rapidly and converges more slowly with $\omega/\omega_c$.  Error in the coefficient $\rho_Q$ spikes around $\omega/\omega_c = 2 \times 10^2$ because the coefficient changes sign there, and small differences in the location of that zero-crossing amount to large relative errors.

Similar results were found for the other isotropic distribution functions we tested -- namely the power-law and kappa distributions -- though for these two no equivalent of the Trubnikov tensor exists, and for the latter there are no fitting formulae either.  Fortunately, the distribution function separates from the numerically difficult portion of the integral (the kernel $\mathbb{K}_{ij}$ in equation \ref{eq:chi_ij_ours}), so significant errors should not arise upon changing distribution functions, so long as they are smooth and well-behaved in $\gamma$.  Comparisons of the absorption coefficients to those from {\tt symphony} for the power-law and kappa distributions agree to within 1\% for the region of parameter space surveyed.

\section{Conclusion}

In this paper we provide a general means for numerically evaluating the susceptibility tensor for arbitrary isotropic distribution functions.   This result can be used to evaluate the modes of a relativistic magnetized plasma, and to find the radiative transfer coefficients related to absorption and Faraday rotation.   We showcase the accuracy and generality of our approach using a series of analytic and numerical tests. The new scheme is implemented in the publicly available code {\tt symphony}, available for free online$^{1}$.

\acknowledgments

This work was supported by a Princeton First Year Fellowship awarded to AP and {\it Quazar Technologies} for MC.  
CFG acknowledges support from NSF grants AST-1333612, AST-1716327,
and PIRE-1743747, and a Romano Professorial Scholarship.

\appendix

\section{Appendix} \label{section:appendix}

In this Appendix we derive our expression for the susceptibility 3-tensor $\chi_{ij}$.   First we provide a brief summary, then a more detailed proof.

Both the standard approach and our approach can be summarized as follows.  First, write down the linearized Vlasov equation for the perturbed distribution $f_1$ around some equilibrium distribution function $f_0$ which is uniform in space and time, in the presence of a uniform magnetic field $\bB$ (equation \ref{eq:Vlasov_eq}).   Find $f_1$ by an integration in time along unperturbed orbits, assuming the perturbed electric field $E_{1,i} \propto \exp(i \bk \cdot \bx - i \omega t)$ (equation \ref{eq:f_1}).  Next, take a first moment of $f_1$ in momentum space to find the current $j_i$ (equation \ref{eq:current}).  Identify this current with $\sigma_{ij}  E_{1,j}$ to find the conductivity tensor and hence $\chi_{ij}$, which is now written as an integral over momentum space and time (equation \ref{eq:stix_45}).  

The standard approach involves rewriting the exponential space and time dependence (which comes from $E_{1,i}$  described in the previous paragraph) in terms of an infinite sum over Bessel functions (equation \ref{eq:phi_integral}), which is then integrated over time (equations \ref{eq:stix_45}-\ref{eq:resonant_denom_integral}).   For a gyrotropic distribution $f_0 = f_0(p_\perp, p_\parallel)$ the result is a two dimensional momentum space integral over an integrand containing the infinite sum and a resonant denominator featuring resonances in both of the two momentum-space integrals (equation \ref{eq:textbook_chi}).  This is the form implicitly used in evaluation of the absorptivities in {\tt symphony}.  Numerical evaluation requires, effectively, evaluation of a three dimensional integral.      


It is worth noting here that Trubnikov (1958; summarized in covariant form by Melrose 1997) carries the calculation a bit further.   Assuming a relativistic thermal (Maxwell-J\"uttner) distribution, he uses a distribution-specific set of manipulations to directly evaluate the momentum space integrals, leaving a single integral over time.

Our approach starts with the standard approach prior to the integration over time: a two dimensional phase space integral, an infinite sum, and an integral over time (equation \ref{eq:stix_45}).   Using Bessel function identities we rewrite the integrand to eliminate the infinite sum (equation \ref{eq:chi_before_cosxi_integration}).   Then we carry out an integral over angle in momentum space.   This last step is restricted to isotropic distribution functions.  The resulting expression (equation \ref{eq:chi_ij_ours}) is a relatively well-behaved two dimensional integral over Lorentz factor and time that is susceptible to numerical evaluation.

\subsection{Standard Approach} \label{section:standard_approach}

In the work below we follow closely Swanson (2003; 2008) and \cite{stix1992}, filling in steps they omit.

The Vlasov equation (in Gaussian units) is
\begin{equation} \label{eq:Vlasov_eq}
\frac{d f}{d t} = \frac{\partial f}{\partial t} + \bvel \cdot \nabla f + q \Big[ \bE + \frac{\bvel}{c} \times \bB \Big] \nabla_p f = 0
\end{equation}
where $f(\bp, t, \bx)$ is the particle distribution function, $q$ is the signed charge (which is negative for electrons), and $\nabla_p f$ is the gradient of $f$ in momentum space.  We are interested in solving the Vlasov equation to linear order in the perturbing field $\bE_1$ for a plasma with a static background magnetic field $\bB_0$ and no background electric field $\bE_0 = 0$.  Formally, we expand the following quantities
\begin{align}
f(\bp, t, \bx) &= f_0(\bp) + f_1(\bp, t, \bx) + ... \\
\bE(t, \bx) &= \bE_1(t, \bx) + ... \\
\bB(t, \bx) &= \bB_0 + \bB_1(t, \bx) + ...
\end{align}
and are interested in solving for $f_1$.  Substituting the above definitions into equation \ref{eq:Vlasov_eq} and dropping terms higher than first order, we have
\begin{equation}
\frac{\partial}{\partial t} (f_0 + f_1) + \bvel \cdot \nabla (f_0 + f_1) + q \Big[ \frac{\bvel}{c} \times \bB_0 \Big] \nabla_p f_1 + q \Big[ \bE_1 + \frac{\bvel}{c} \times \bB_1 \Big] \nabla_p f_0 = 0.
\end{equation}
Since $f_0$ is assumed to be independent of time and position, $\partial f_0 / \partial t = 0$ and $\nabla f_0 = 0$.  Using this fact and rearranging
\begin{equation} \label{eq:vlasov_step_2}
\frac{\partial f_1}{\partial t} + \bvel \cdot \nabla f_1 + q \Big[ \frac{\bvel}{c} \times \bB_0 \Big] \nabla_p f_1 = - q \Big[ \bE_1 + \frac{\bvel}{c} \times \bB_1 \Big] \nabla_p f_0.
\end{equation}
Note that the left hand side of the equation is equal to the Vlasov equation for a distribution $f_1(\bp, t, \bx)$ for a particle only under the influence of the static background magnetic field, $\bB_0$.  The trajectory of the particle only under the influence of $\bB_0$ is conventionally called its \textit{unperturbed orbit}, for which the following approach is named.  Thus using equation \ref{eq:Vlasov_eq} we can rewrite equation \ref{eq:vlasov_step_2} as
\begin{equation}
\frac{d f_1}{d t} \bigg|_{\mathrm{unperturbed~orbit}} = - q \Big[ \bE_1 + \frac{\bvel}{c} \times \bB_1 \Big] \nabla_p f_0,
\end{equation}
which can be expressed as an integral
\begin{equation} \label{eq:f1_integral_form}
f_1(\bp, t, \bx)  = - q \int_{-\infty}^{t} dt' \Big[ \bE_1(t', \bx') + \frac{\bvel'}{c} \times \bB_1(t', \bx') \Big] \nabla_{p'} f_0,
\end{equation}
where the integral is taken over the aforementioned unperturbed particle orbit, denoted by primed variables $t'$, $\bvel'$, and $\bp'$.  This integral is taken over the entire history of the particle along its unperturbed orbit, from $t' = -\infty$ to when the perturbing field is applied at $t' = t$.

We can assume that the perturbing electric and magnetic fields are of the form
\begin{align}
\bE_1(t, \bx) &= \bE_{c} e^{-i(\omega t - \bk \cdot \bx)} \\
\bB_1(t, \bx) &= \bB_{c} e^{-i(\omega t - \bk \cdot \bx)}
\end{align}
where $\bE_c$ and $\bB_c$ are the constant vector amplitudes of the electric and magnetic fields, and $\mathrm{Im}(\omega) > 0$ (see \S \ref{section:review_defns} for a discussion).  Using this assumption and Maxwell's equations we can rewrite $\bB_1$ as
\begin{equation}
\bB_1 = \frac{c}{\omega} \bk \times \bE_1 = \frac{c}{\omega} \bk \times \bE_c e^{-i(\omega t - \bk \cdot \bx)};
\end{equation}
substituting this equation into equation \ref{eq:f1_integral_form} yields
\begin{equation}
f_1(\bp, t, \bx)  = - q \int_{-\infty}^{t} dt' e^{-i(\omega t' - \bk \cdot \bx')} \Big[ \bE_c + \frac{\bvel'}{c} \times \frac{c}{\omega} \bk \times \bE_c \Big] \nabla_{p'} f_0,
\end{equation}
which, after expanding the vector triple product, can be written 
\begin{equation} \label{eq:f_1}
f_1(\bp, t, \bx)  = - q \int_{-\infty}^{t} dt' e^{-i(\omega t' - \bk \cdot \bx')} \bE_c \bigg[1 + \frac{\cdot \bvel' \bk - \bvel' \cdot \bk}{\omega} \bigg] \cdot \nabla_{p'} f_0.
\end{equation}
The odd notation in the above equation -- with the vector quantity $\bE_c$ pulled out of the brackets, splitting the dot product in the $\cdot \bvel' \bk$ term -- is kept for easy comparison with equation 33 of \cite{stix1992}.

\subsubsection{Solving for the Unperturbed Orbits}

Now we need to express $\bvel'$ and $\bx'$ in terms of $t'$.  We can do this by noting that a particle on the unperturbed trajectory $\bx'(t')$ is described by:
\begin{equation}
\textbf{F} = \gamma m \frac{d \bvel'}{d t'} = \frac{q \bvel'}{c} \times \bB_0
\end{equation}
which is also subject to the constraint that at $t' = t$ we must have $\bvel' = \bvel$.  Note that the acceleration is always perpendicular to the velocity; as a result, $|v|$ will remain constant along the entirety of the unperturbed orbit, up to $t' = t$ where $\bvel' = \bvel$.  Thus we may use $v$ in the definition of $\gamma$ rather than $v'$, making $\gamma$ a constant in the differential equation.

The static magnetic field is conventionally taken to be parallel to the $\hat{z}$ axis.  It will be easier to solve for unknown coefficients if we define $\tau' \equiv t - t'$, note that $d\tau' = - dt'$, and rewrite this differential equation
\begin{equation}
\gamma m \frac{d \bvel'}{d \tau'} = - \frac{q \bvel'}{c} \times B_0 \hat{z}.
\end{equation}
Breaking this vector equation into components and defining the (signed) nonrelativistic cyclotron frequency $\omega_c = q B /(m c)$ as well as the (signed) relativistic version $\Omega_c \equiv \omega_c / \gamma$, one may solve the equations to find
\begin{align}
v_x' &= - v_y \sin(\Omega_c \tau') + v_x \cos(\Omega_c \tau') \\
v_y' &= v_x \sin(\Omega_c \tau') + v_y \cos(\Omega_c \tau') \\
v_z' &= v_z.
\end{align}
Integrating with respect to $dt' = -d\tau'$ and applying the boundary condition $\bx' = \bx$ at $\tau' = 0$ results in the particle's full trajectory
\begin{align}
x' &= \frac{- v_y}{\Omega_c} \cos(\Omega_c \tau') + \frac{- v_x}{\Omega_c} \sin(\Omega_c \tau') + x + \frac{v_y}{\Omega_c} \\
y' &= \frac{v_x}{\Omega_c} \cos(\Omega_c \tau') + \frac{- v_y}{\Omega_c} \sin(\Omega_c \tau') + y - \frac{v_x}{\Omega_c} \\
z' &= - v_z \tau' + z.
\end{align}

\subsubsection{Integrating Over Unperturbed Orbits}

We can now substitute our values for $\bvel'$ and $\bx'$ into equation \ref{eq:f_1} (reproduced below) and continue our simplification of the integral
\begin{equation} \tag{\ref{eq:f_1}}
f_1(\bvel, t, \bx)  = - q \int_{-\infty}^{t} dt' e^{-i(\omega t' - \bk \cdot \bx')} \bE_c \bigg[1 + \frac{\cdot \bvel' \bk - \bvel' \cdot \bk}{\omega} \bigg] \cdot \nabla_{p'} f_0.
\end{equation}
We can immediately see that we need to rewrite the exponential in terms of $\tau'$ using the unperturbed orbit $\bx'(\tau')$.  Doing so, we find
\begin{multline} \label{eq:exponential_factor}
-i(\omega t' - \bk \cdot \bx') = -i(\omega t - \bk \cdot \bx) + \frac{i v_x}{ \Omega_c} [-k_x \sin( \Omega_c \tau') - k_y (1 - \cos( \Omega_c \tau'))] 
\\ + \frac{i v_y}{ \Omega_c} [-k_y \sin( \Omega_c \tau') + k_x (1 - \cos( \Omega_c \tau'))] + i(\omega - k_z v_z)\tau'.
\end{multline}

In order to progress further we must make the assumption that $f_0$ is gyrotropic, meaning it is independent of the gyrophase $\phi$.  This assumption is equivalent to supposing that $f_0 = f_0 (p_\perp, p_\parallel)$, where $p_\perp$ is the component of the momentum perpendicular to the magnetic field $\bB = B \hat{z}$, and $p_\parallel$ is the component parallel to $\bB$.  Analogous definitions are made for the perpendicular and parallel velocities, which appear below.

If we first introduce the notation
\begin{align}
\frac{\partial f_0}{\partial p'_\perp} &\equiv f_{0\perp} \\
\frac{\partial f_0}{\partial p'_z}     &\equiv f_{0z}
\end{align}
we can write
\begin{align}
\frac{\partial f_0}{\partial p_x'} &= \frac{v_x'}{v_\perp} f_{0\perp} \\
\frac{\partial f_0}{\partial p_y'} &= \frac{v_y'}{v_\perp} f_{0\perp} \\
\frac{\partial f_0}{\partial p_z'} &= f_{0z},
\end{align}
and we can now expand the other factor from equation \ref{eq:f_1}.  Doing so yields
\begin{multline}
\bE_c \bigg[1 + \frac{\cdot \bvel' \bk - \bvel' \cdot \bk}{\omega} \bigg] \cdot \nabla_{p'} f_0 = \\
(E_x v_x' + E_y v_y') \bigg[\frac{f_{0\perp}}{v_\perp} + \frac{k_z}{\omega} \bigg(f_{0z} - \frac{v_z'}{v_\perp} f_{0\perp} \bigg) \bigg] \\ 
+ E_z \bigg[ f_{0z} - \frac{k_x v_x' + k_y v_y'}{\omega} \bigg( f_{0z} - \frac{v_z'}{v_\perp} f_{0\perp} \bigg) \bigg].
\end{multline}
Substituting the values for $v_x'$, $v_y'$, and $v_z'$ gives the final result for this factor
\begin{multline} \label{eq:Stix_factor_simplified}
\bE_c \bigg[1 + \frac{\cdot \bvel' \bk - \bvel' \cdot \bk}{\omega} \bigg] \cdot \nabla_{p'} f_0 = \\
(v_x \cos( \Omega_c \tau') - v_y \sin( \Omega_c \tau')) \bigg[ \frac{E_x f_{0\perp}}{v_\perp} + \frac{E_x k_z - E_z k_x}{\omega} \bigg( f_{0z} - \frac{v_z}{v_\perp} f_{0\perp} \bigg) \bigg] \\
+ (v_x \sin( \Omega_c \tau') + v_y \cos( \Omega_c \tau')) \bigg[ \frac{E_y f_{0\perp}}{v_\perp} + \frac{E_y k_z - E_z k_y}{\omega} \bigg( f_{0z} - \frac{v_z}{v_\perp} f_{0\perp} \bigg) \bigg] \\
+ E_z f_{0z}.
\end{multline}

We can now work toward evaluating the integral in equation \ref{eq:f_1}.  The complicated exponential factor in equation \ref{eq:exponential_factor} can be simplified by defining
\begin{align}
a &= \omega - k_z v_z\\
b &= \frac{k_\perp v_\perp}{ \Omega_c} \label{eq:b}
\end{align}
so that equation \ref{eq:exponential_factor} (after being exponentiated) can be written
\begin{equation} \label{eq:simplified_exponential}
e^{-i(\omega t' - \bk \cdot \bx')} = e^{-i(\omega t - \bk \cdot \bx)} e^{-i b\sin(\phi - \psi +  \Omega_c \tau')} e^{ + i b \sin(\phi - \psi)} e^{ i a \tau' }.
\end{equation}

At this point we may write $f_1$ explicitly using equations \ref{eq:Stix_factor_simplified} and \ref{eq:simplified_exponential} as
\begin{multline}
f_1(\bvel, t, \bx)  = - q \int_{-\infty}^{t} dt' e^{-i(\omega t - \bk \cdot \bx)} e^{-i b\sin(\phi - \psi +  \Omega_c \tau')} e^{ + i b \sin(\phi - \psi)} e^{ i a \tau' } \\
\times \Bigg\{ (v_x \cos( \Omega_c \tau') - v_y \sin( \Omega_c \tau')) \bigg[ \frac{E_x f_{0\perp}}{v_\perp} + \frac{E_x k_z - E_z k_x}{\omega} \bigg( f_{0z} - \frac{v_z}{v_\perp} f_{0\perp} \bigg) \bigg] \\
+ (v_x \sin( \Omega_c \tau') + v_y \cos( \Omega_c \tau')) \bigg[ \frac{E_y f_{0\perp}}{v_\perp} + \frac{E_y k_z - E_z k_y}{\omega} \bigg( f_{0z} - \frac{v_z}{v_\perp} f_{0\perp} \bigg) \bigg] + E_z f_{0z} \Bigg\}.
\end{multline}
Using this expression, changing the integral from one over $t'$ to one over $\tau'$ ($\in (0, \infty)$), suppressing the $\exp(-i(\omega t - \bk \cdot \bx))$ and interpreting the integral as the Fourier amplitude of the distribution function,
\begin{multline} \label{eq:f_1_explicit}
f_1(\bvel, \omega, \bk)  = q \int_{0}^{\infty} d\tau' e^{-i b\sin(\phi - \psi +  \Omega_c \tau')} e^{ + i b \sin(\phi - \psi)} e^{ i a \tau' } \\
\times \Bigg\{ (v_x \cos( \Omega_c \tau') - v_y \sin( \Omega_c \tau')) \bigg[ \frac{E_x f_{0\perp}}{v_\perp} + \frac{E_x k_z - E_z k_x}{\omega} \bigg( f_{0z} - \frac{v_z}{v_\perp} f_{0\perp} \bigg) \bigg] \\
+ (v_x \sin( \Omega_c \tau') + v_y \cos( \Omega_c \tau')) \bigg[ \frac{E_y f_{0\perp}}{v_\perp} + \frac{E_y k_z - E_z k_y}{\omega} \bigg( f_{0z} - \frac{v_z}{v_\perp} f_{0\perp} \bigg) \bigg] + E_z f_{0z} \Bigg\} .
\end{multline}

\subsubsection{Finding the Conductivity Tensor $\sigma_{ij}$}

Equation \ref{eq:f_1_explicit} can be used to find the current
\begin{equation} \label{eq:current}
j_i(\bp, \omega, \bk) = q \int d^3p~v_i f_1(\bp, \omega, \bk) = \sigma_{ij} E_j.
\end{equation}
Inserting $f_1$, we find 
\begin{multline}
j_i = q^2 \int d^3p~v_i \int_{0}^{\infty} d\tau' e^{-i b\sin(\phi - \psi +  \Omega_c \tau')} e^{ + i b \sin(\phi - \psi)} e^{ i a \tau' } \\
\times \Bigg\{ (v_x \cos( \Omega_c \tau') - v_y \sin( \Omega_c \tau')) \bigg[ \frac{E_x f_{0\perp}}{v_\perp} + \frac{E_x k_z - E_z k_x}{\omega} \bigg( f_{0z} - \frac{v_z}{v_\perp} f_{0\perp} \bigg) \bigg] \\
+ (v_x \sin( \Omega_c \tau') + v_y \cos( \Omega_c \tau')) \bigg[ \frac{E_y f_{0\perp}}{v_\perp} + \frac{E_y k_z - E_z k_y}{\omega} \bigg( f_{0z} - \frac{v_z}{v_\perp} f_{0\perp} \bigg) \bigg] + E_z f_{0z} \Bigg\} ;
\end{multline}
at this point many authors elect to evaluate the $\tau'$ integral; we instead follow the treatment of \cite{stix1992}, who first switches to cylindrical coordinates $\{p_\perp, p_z, \phi\}$ and evaluates the angular momentum-space integral over $\phi$.

To begin, we express the components of $\bvel$ and $\bk$ in cylindrical coordinates
\begin{align}
v_x &= v_\perp \cos(\phi) \label{eq:v_x} \\
v_y &= v_\perp \sin(\phi) \label{eq:v_y} \\
k_x &= k_\perp \cos(\psi) \label{eq:k_x} \\
k_y &= k_\perp \sin(\psi) \label{eq:k_y},
\end{align}
where we have introduced the polar angle $\psi$ to denote the angle of the wavevector in the $x$-$y$ plane, and the components $v_z$ and $k_z$ are identical to their Cartesian counterparts.  
Stix immediately simplifies the computation by fixing coordinates such that $\psi = 0$, resulting in $k_x = k_\perp$, $k_y = 0$, and $\mathrm{Re}(k_x) > 0$. This choice does not amount to a loss of generality, as one may simply rotate the resultant susceptibility tensor at the end of the computation to return the $\psi \neq 0$ case.

Substituting the new definitions for $\bvel$ and $\bk$, it is now possible to evaluate the $\phi$ integral using the following known integrals:
\begin{equation} \label{eq:phi_integral}
\int_{0}^{2 \pi} d\phi e^{-ib[\sin(\phi + \Omega_c \tau') - \sin \phi]}
\begin{pmatrix}
\sin \phi \sin(\phi + \Omega_c \tau') \\
\sin \phi \cos(\phi + \Omega_c \tau') \\
\cos \phi \sin(\phi + \Omega_c \tau') \\
\cos \phi \cos(\phi + \Omega_c \tau') \\
1 \\
\sin \phi \\
\cos \phi \\
\sin(\phi + \Omega_c \tau') \\
\cos(\phi + \Omega_c \tau')
\end{pmatrix}
= 2 \pi \sum_{n = -\infty}^{\infty} e^{-i n \Omega_c \tau'}
\begin{pmatrix}
(J_n')^2 \\
- \frac{i n}{b} J_n J_n' \\
\frac{i n}{b} J_n J_n' \\
\frac{n^2}{b^2} J_n^2 \\
J_n^2 \\
-i J_n J_n' \\
\frac{n}{b} J_n^2 \\
i J_n J_n' \\
\frac{n}{b} J_n^2
\end{pmatrix},
\end{equation}
which are all derived from Bessel function orthogonality relations.  The arguments of the Bessel functions in equation \ref{eq:phi_integral} are all $b$, given in equation \ref{eq:b}.  Computing the integral and arranging the terms into a matrix, we find:
\begin{multline}
\bj = -\frac{q^2}{m} \int_{-\infty}^{\infty} \int_{0}^{\infty} v_\perp d v_\perp d v_z \int_{0}^{\infty} d\tau' e^{ i a \tau' } 2 \pi \sum_{n = -\infty}^{\infty} e^{- i n \epsilon \Omega_c \tau'} \\
\begin{pmatrix}
v_\perp \frac{n^2}{b^2} J_n^2 U & v_\perp \frac{i n}{b} J_n J_n' U & \frac{n}{b} J_n^2 v_\perp W \\
-v_\perp \frac{i n}{b} J_n J_n' U & v_\perp (J_n')^2 U & - i J_n J_n' v_\perp W \\
v_z \frac{n}{b} J_n^2 U & v_z i J_n J_n' U & J_n^2 v_z W
\end{pmatrix}
\cdot
\begin{pmatrix}
E_x \\
E_y \\
E_z
\end{pmatrix}
= \sigma \bE,
\end{multline}
where
\begin{align} 
U &\equiv \frac{\partial f_0}{\partial p_\perp} + \frac{k_z}{\omega} \bigg( v_\perp \frac{\partial f_0}{\partial p_z} - v_z \frac{\partial f_0}{\partial p_\perp} \bigg) \\
W &\equiv \Big( 1 - \frac{n \Omega_c}{\omega} \Big) \frac{\partial f_0}{\partial p_z} + \frac{n \Omega_c p_z}{p_\perp} \frac{\partial f_0}{\partial p_\perp}
\end{align}
following Stix.  Hence one may read off the conductivity tensor $\sigma$ and then use equations \ref{eq:K_and_sigma}-\ref{eq:K_and_chi} to find the susceptibility tensor.

\subsubsection{Arriving at the Standard Form of the Susceptibility Tensor}

Computing the susceptibility tensor from the conductivity tensor yields
\begin{multline} \label{eq:stix_45}
\chi_{ij} = -\frac{i q^2}{\omega \varepsilon_0} \int_{-\infty}^{\infty} \int_{0}^{\infty} \frac{p_\perp}{\gamma} d p_\perp d p_z \int_{0}^{\infty} d\tau' e^{ i a \tau' } 2 \pi \sum_{n = -\infty}^{\infty} e^{- i n  \Omega_c \tau'} \\
\begin{pmatrix}
p_\perp \frac{n^2}{b^2} J_n^2 U & p_\perp \frac{i n}{b} J_n J_n' U & \frac{n}{b} J_n^2 p_\perp W \\
-p_\perp \frac{i n}{b} J_n J_n' U & p_\perp (J_n')^2 U & - i J_n J_n' p_\perp W \\
p_z \frac{n}{b} J_n^2 U & p_z i J_n J_n' U & J_n^2 p_z W
\end{pmatrix}.
\end{multline}

At this point all authors use the fact that $\mathrm{Im}(\omega) > 0$ to evaluate the $\tau'$ integral in equation \ref{eq:stix_45}, resulting in the following (\cite{stix1992} Ch. 10, eq. 44):
\begin{equation} \label{eq:resonant_denom_integral}
-q \int_0^\infty d \tau' \exp[i (\omega - k_z v_z - n  \Omega_c) \tau'] = \frac{-i q}{\omega - k_z v_z - n  \Omega_c},
\end{equation}
where the right-hand side involves a resonant denominator that complicates both the $p_\perp$ and $p_z$ integrals significantly.  We choose not to evaluate this integral at this time.  Instead, we have developed a novel approach that involves analytically computing the infinite sum, then changing variables so that one of the two remaining momentum-space integrals may also be done analytically.  For completeness, we include the final standard form of the susceptibility tensor (\cite{stix1992}):
\begin{multline} \label{eq:textbook_chi}
\chi_{ij} = \frac{\omega_p^2}{\omega} \sum_{n = -\infty}^{\infty} \int_{0}^{\infty} \frac{2 \pi p_\perp}{\gamma} dp_\perp \int_{-\infty}^{\infty} dp_z \frac{1}{\omega - k_z p_z - n \Omega_c} \\
\begin{pmatrix}
p_\perp \frac{n^2}{b^2} J_n^2 U & p_\perp \frac{i n}{b} J_n J_n' U & \frac{n}{b} J_n^2 p_\perp W \\
-p_\perp \frac{i n}{b} J_n J_n' U & p_\perp (J_n')^2 U & - i J_n J_n' p_\perp W \\
p_z \frac{n}{b} J_n^2 U & p_z i J_n J_n' U & J_n^2 p_z W
\end{pmatrix}.
\end{multline}

\subsection{Novel Integration Method} \label{section:novel_int_method}

Now we present our novel approach for integrating the susceptibility tensor.  We begin with equation \ref{eq:stix_45} and change variables from $p_\perp, p_z$ to $\gamma \equiv \sqrt{1 + (p_\perp/mc)^2 + (p_z/mc)^2}$; $\cos \xi$, defined with particle momentum pitch angle $\xi \equiv \mathrm{arctan2}(p_z, p_\perp)$ (where $\mathrm{arctan2}$ is the two-argument arctangent); $\tau = \mathrm{Re}(\omega) \tau' / \gamma$.  Substituting these new definitions yields
\begin{multline}
\chi_{ij}(\omega, \bk)  = -\frac{i q^2}{\varepsilon_0 m \omega \mathrm{Re}(\omega)} \int_{1}^{\infty} \int_{-1}^{1} (m c)^3 \gamma^3 \beta^2 d\gamma d \cos \xi \int_{0}^{\infty} d\tau e^{ i \gamma \frac{\omega}{\mathrm{Re}(\omega)} \tau } e^{- i \alpha \cos \xi} \\
\times 2 \pi \sum_{n = -\infty}^{\infty} e^{- \frac{i n \omega_c \tau}{\mathrm{Re}(\omega)} }
\begin{pmatrix}
(m c) \sin \xi \frac{n^2}{b^2} J_n^2 U & (m c) \sin \xi \frac{i n}{b} J_n J_n' U & (m c) \frac{n}{b} J_n^2 \sin \xi W \\
-(m c) \sin \xi \frac{i n}{b} J_n J_n' U & (m c) \sin \xi (J_n')^2 U & - (m c) i J_n J_n' \sin \xi W \\
(m c) \cos \xi \frac{n}{b} J_n^2 U & (m c) \cos \xi i J_n J_n' U & (m c) J_n^2 \cos \xi W
\end{pmatrix},
\end{multline}
where 
\begin{equation} \tag{\ref{eq:alpha_defn}}
\alpha = \frac{\gamma \beta c k \cos (\theta)}{\mathrm{Re}(\omega)} \tau.
\end{equation}

The infinite sums at this point may all be evaluated analytically if one makes use of a couple of well-known Bessel function identities and the Graf Addition Theorem
\begin{align}
n J_n(z) &= \frac{z}{2} (J_{n-1}(z) + J_{n+1}(z)) \\
J_n'(z)  &= \frac{1}{2} (J_{n-1}(z) - J_{n+1}(z)). \\
\sum_{n = - \infty}^{\infty} e^{i n \theta} J_{n + \nu}(x) J_n(y) &= \bigg( \frac{x - y e^{-i \theta}}{x - y e^{i \theta}} \bigg)^{\nu/2} J_\nu (\sqrt{x^2 + y^2 - 2 x y \cos (\theta)}).
\end{align}
The specific cases of these formulae applied to the sums in the susceptibility tensor are worked out and provided in \S \ref{section:bessel_sums}.  Applying these formulae yields
\begin{multline} \label{eq:chi_before_cosxi_integration}
\chi_{ij}(\omega, \bk)  = -\frac{2 \pi i q^2}{\varepsilon_0 m \omega \mathrm{Re}(\omega)} \int_{1}^{\infty} \int_{-1}^{1} (m c)^4 \gamma^3 \beta^2 d\gamma d \cos \xi \int_{0}^{\infty} d\tau e^{ i \gamma \frac{\omega}{\mathrm{Re}(\omega)} \tau} e^{- i \alpha \cos \xi} \\
\times
\begin{pmatrix}
\frac{1}{2} (\sin \xi)^2 \bigg[ C J_0 - J_2 \bigg] U & \frac{1}{2} (\sin \xi)^2 S J_0 U & i (\sin \xi) (\cos \xi) C_2 J_1 W \\
-\frac{1}{2} (\sin \xi)^2 S J_0 U & \frac{1}{2} (\sin \xi)^2 \bigg[ C J_0 + J_2 \bigg] U & - i (\sin \xi) (\cos \xi) S_2 J_1 W \\
i (\sin \xi) (\cos \xi) C_2 J_1 U & i (\sin \xi) (\cos \xi) S_2 J_1 U & (\cos \xi)^2 J_0 W
\end{pmatrix}
\end{multline}
where $C \equiv \cos( \frac{ \omega_c \tau}{\mathrm{Re}(\omega)} )$, $C_2 \equiv \cos(\frac{ \omega_c \tau }{2 \mathrm{Re}(\omega)} )$, $S \equiv \sin(\frac{ \omega_c \tau}{\mathrm{Re}(\omega)} )$, and $S_2 \equiv \sin(\frac{ \omega_c \tau}{2 \mathrm{Re}(\omega)} )$.  The arguments of the Bessel functions are $\delta \sin \xi$ and are omitted to save space.  The definition of $\delta$ is reproduced here for convenience: 
\begin{equation} \tag{\ref{eq:delta_defn}}
\delta = \frac{2 \gamma \beta c k \sin (\theta)}{\omega_c} \sin \Big( \frac{\omega_c}{2 \mathrm{Re}(\omega)} \tau \Big).
\end{equation}

The $\cos \xi$ integral in equation \ref{eq:chi_before_cosxi_integration} above may be evaluated analytically (see \S \ref{section:cos_xi_integrals}) for each of the susceptibility tensor components, as long as the distribution function is independent of $\xi$ (in other words, as long as $f$ is isotropic).  

Evaluating these integrals, we find
\begin{align}
\mathcal{I}_1(0) &= \frac{2 ((2 \alpha^2 + (\alpha^2 - 1) \delta^2 + \delta^4) \sin A - (2 \alpha^2 - \delta^2) A \cos A)}{A^5} \tag{\ref{eq:I_1_of_0}}\\
\mathcal{I}_1(2) &= - \frac{2 \delta^2 (3 A \cos A + (A^2 - 3) \sin A)}{A^5} \tag{\ref{eq:I_1_of_2}}\\
\mathcal{I}_2(1) &= \frac{ 2 i \alpha \delta \big( 3 A \cos A + (A^2 - 3) \sin A \big)}{A^5} \tag{\ref{eq:I_2_of_1}}\\
\mathcal{I}_3(0) &= \frac{6 \alpha^2 \cos A}{A^4} - \frac{2 \cos A}{A^2} + \frac{6 \delta^2 \sin A}{A^5} - \frac{4 \sin A}{A^3} + \frac{2 \alpha^2 \sin A}{A^3}, \tag{\ref{eq:I_3_of_0}}
\end{align}
where $A = \sqrt{\alpha^2 + \delta^2}$.  All of the above results have been checked numerically, and the derivations are shown in \S \ref{section:cos_xi_integrals}.

The final susceptibility tensor is shown in equation \ref{eq:chi_ij_ours}.

\subsection{Analytic Evaluation of Bessel Function Sums} \label{section:bessel_sums}

Evaluating the required sums yields
\begin{equation} \label{eq:first_bessel_sum}
\sum_{n=-\infty}^{\infty} e^{- \frac{i n  \omega_c \tau}{\mathrm{Re}(\omega)}} n^2 J_n^2 = 
\frac{b^2}{2} \bigg[ \cos \Big(  \frac{\omega_c \tau}{\mathrm{Re}(\omega)} \Big) J_0 \bigg( 2 b \sin \bigg( \frac{ \omega_c \tau}{2 \mathrm{Re}(\omega)} \bigg) \bigg) - J_2 \bigg( 2 b \sin \bigg( \frac{ \omega_c \tau}{2 \mathrm{Re}(\omega)} \bigg) \bigg) \bigg]
\end{equation}
\begin{equation}
\sum_{n=-\infty}^{\infty} e^{- \frac{i n  \omega_c \tau}{\mathrm{Re}(\omega)}} n J_n^2 = -i b \cos \bigg( \frac{ \omega_c \tau}{2 \mathrm{Re}(\omega)} \bigg) J_1 \bigg( 2 b \sin \bigg( \frac{ \omega_c \tau}{2 \mathrm{Re}(\omega)} \bigg) \bigg)
\end{equation}
\begin{equation}
\sum_{n=-\infty}^{\infty} e^{- \frac{i n  \omega_c \tau}{\mathrm{Re}(\omega)}} J_n^2 = J_0 \bigg( 2 b \sin \bigg( \frac{ \omega_c \tau}{2 \mathrm{Re}(\omega)} \bigg) \bigg)
\end{equation}
\begin{equation}
\sum_{n=-\infty}^{\infty} e^{- \frac{i n  \omega_c \tau}{\mathrm{Re}(\omega)}} n J_n J_n' = - \frac{i b}{2} \sin \Big( \frac{\omega_c \tau}{\mathrm{Re}(\omega)} \Big) J_0 \bigg( 2 b \sin \bigg( \frac{ \omega_c \tau}{2 \mathrm{Re}(\omega)} \bigg) \bigg)
\end{equation}
\begin{equation}
\sum_{n=-\infty}^{\infty} e^{- \frac{i n  \omega_c \tau}{\mathrm{Re}(\omega)}} J_n J_n' = - \sin \bigg( \frac{ \omega_c \tau}{2 \mathrm{Re}(\omega)} \bigg) J_1 \bigg( 2 b \sin \bigg( \frac{ \omega_c \tau}{2 \mathrm{Re}(\omega)} \bigg) \bigg)
\end{equation}
\begin{equation} \label{eq:last_bessel_sum}
\sum_{n=-\infty}^{\infty} e^{- \frac{i n  \omega_c \tau}{\mathrm{Re}(\omega)}} (J_n')^2 = \frac{1}{2} \bigg[ \cos \Big(  \frac{\omega_c \tau}{\mathrm{Re}(\omega)} \Big) J_0 \bigg( 2 b \sin \bigg( \frac{ \omega_c \tau}{2 \mathrm{Re}(\omega)} \bigg) \bigg) + J_2 \bigg( 2 b \sin \bigg( \frac{ \omega_c \tau}{2 \mathrm{Re}(\omega)} \bigg) \bigg) \bigg],
\end{equation}
where the arguments of the Bessel functions on the left-hand sides of equations \ref{eq:first_bessel_sum}-\ref{eq:last_bessel_sum} are $b$, as given in equation \ref{eq:b}.

\subsection{Analytic Integration of $\cos \xi$ Integrals} \label{section:cos_xi_integrals}

All of the $\cos \xi$ integrals in equation \ref{eq:chi_before_cosxi_integration} are one of three types:
\begin{align}
\mathcal{I}_1(n) &\equiv \int_{-1}^{1} \Big( \sqrt{1 - x^2} \Big)^2 e^{- i \alpha x} J_n \Big( \delta \sqrt{1 - x^2} \Big) dx  \\
\mathcal{I}_2(n) &\equiv \int_{-1}^{1} x \sqrt{1 - x^2} e^{- i \alpha x} J_n \Big( \delta \sqrt{1 - x^2} \Big) dx  \\
\mathcal{I}_3(n) &\equiv \int_{-1}^{1} x^2 e^{- i \alpha x} J_n \Big( \delta \sqrt{1 - x^2} \Big) dx.
\end{align}
The integrals $\mathcal{I}_1$ and $\mathcal{I}_3$ can be expressed in terms of the integral 
\begin{equation}
\mathcal{I}^*(n) = \int_{0}^{1} \cos(\alpha x) J_n \Big( \delta \sqrt{1 - x^2} \Big) dx,
\end{equation}
which can be evaluated analytically; $\mathcal{I}_2$ must be handled separately, but it can also be evaluated analytically via a similar method.

\subsubsection{$\mathcal{I}^*(n)$}

Now we must evaluate $\mathcal{I}^*(n)$.  It can be done by manipulating integral number 6.727 from \cite{gradshteyn2007}:
\begin{multline}
\mathcal{I}_{GR}(n) \equiv \int_{0}^{1} \frac{\cos(\alpha x)}{\sqrt{1 - x^2}} J_n \Big( \delta \sqrt{1 - x^2} \Big) dx = \\
\frac{\pi}{2} J_{\frac{n}{2}} \Big( \frac{1}{2} \Big( \sqrt{\alpha^2 + \delta^2} - \alpha \Big) \Big) J_{\frac{n}{2}} \Big( \frac{1}{2} \Big( \sqrt{\alpha^2 + \delta^2} + \alpha \Big) \Big),
\end{multline}
which is erroneously listed as only valid for $\alpha, \delta$ real (compare to equation 10.9.27 of \cite{olver2018}, which allows for complex $\alpha, \delta$ following the substitution $z, \zeta = \frac{1}{2}(A \pm \alpha)$ and change of measure $x \equiv \cos \theta$).  Looking back at $\mathcal{I}^*(n)$, we can integrate it by parts, choosing $\cos(\alpha x)$ to be $dv$ and $J_n(\delta \sqrt{1- x^2})$ to be $u$; this procedure yields
\begin{multline}
\mathcal{I}^*(n) = \bigg[ \frac{1}{\alpha} \sin(\alpha x) J_n \Big( \delta \sqrt{1 - x^2} \Big) \bigg]_{0}^{1} \\
- \int_{0}^{1} \frac{1}{\alpha} \sin(\alpha x) \frac{\delta}{2} \frac{x}{\sqrt{1 - x^2}} \bigg[ J_{n+1} \Big( \delta \sqrt{1 - x^2} \Big) - J_{n-1} \Big( \delta \sqrt{1 - x^2} \Big) \bigg] dx
\end{multline}
which can be rewritten as
\begin{multline}
\mathcal{I}^*(n) = \bigg[ \frac{1}{\alpha} \sin(\alpha x) J_n \Big( \delta \sqrt{1 - x^2} \Big) \bigg]_{0}^{1} - \frac{\delta}{2 \alpha} \bigg[ \int_{0}^{1} x \frac{\sin(\alpha x)}{\sqrt{1 - x^2}} J_{n+1} \Big( \delta \sqrt{1 - x^2} \Big) dx \\
- \int_{0}^{1} x \frac{\sin(\alpha x)}{\sqrt{1 - x^2}} \bigg[ J_{n-1} \Big( \delta \sqrt{1 - x^2} \Big) dx \bigg]\bigg].
\end{multline}
Rearranging, we find
\begin{equation}
\mathcal{I}^*(n) = \bigg[ \frac{1}{\alpha} \sin(\alpha x) J_n \Big( \delta \sqrt{1 - x^2} \Big) \bigg]_{0}^{1} + \frac{\delta}{2 \alpha} \bigg[ \frac{\partial}{\partial \alpha} \mathcal{I}_{GR}(n+1) - \frac{\partial}{\partial \alpha} \mathcal{I}_{GR}(n-1)\bigg].
\end{equation}

\subsubsection{$\mathcal{I}_1(n)$}

The integral $\mathcal{I}_1$ can be written in terms of $\mathcal{I}^*$ by first using Euler's formula and writing the exponentials as $\cos(-\alpha x) + i \sin(-\alpha x)$.  The other terms in both integrands are all even, and the integral is over the symmetric interval $[-1, 1]$, so the imaginary parts of the integrals must be zero by symmetry.  Now we can rewrite $\mathcal{I}_1$ as
\begin{equation}
\mathcal{I}_1(n) = 2 \int_{0}^{1} (1 - x^2) \cos(\alpha x) J_n \Big( \delta \sqrt{1 - x^2} \Big) dx,
\end{equation}
which can be written as
\begin{equation}
\mathcal{I}_1(n) = 2 \mathcal{I}^*(n) + 2 \frac{\partial^2 \mathcal{I}^*(n)}{\partial \alpha^2}.
\end{equation}

This integral appears for two different values of $n$: $0$ and $2$.  The former results in
\begin{equation}
\mathcal{I}_1(0) = \frac{2 ((2 \alpha^2 + (\alpha^2 - 1) \delta^2 + \delta^4) \sin A - (2 \alpha^2 - \delta^2) A \cos A)}{A^5}
\end{equation}
and the latter results in
\begin{equation}
\mathcal{I}_1(2) = - \frac{2 \delta^2 (3 A \cos A + (A^2 - 3) \sin A)}{A^5}.
\end{equation}

\subsubsection{$\mathcal{I}_2(n)$}

The integrand of $\mathcal{I}_2(n)$ is odd, so the symmetry of the integration interval now picks out the imaginary part and the real part is zero:
\begin{equation}
\mathcal{I}_2(n) = -2 i \int_{0}^{1} x \sqrt{1 - x^2} \sin(\alpha x) J_n(\delta \sqrt{1 - x^2}) dx.
\end{equation}
Multiplying by $1 = \sqrt{1 - x^2}/\sqrt{1 - x^2}$ yields
\begin{equation}
\mathcal{I}_2(n) = -2 i \int_{0}^{1} x (1 - x^2) \frac{\sin(\alpha x)}{\sqrt{1 - x^2}} J_n(\delta \sqrt{1 - x^2}) dx = i \frac{\partial \mathcal{I}_1(n)}{\partial \alpha}
\end{equation}
which can be written as
\begin{equation}
\mathcal{I}_2(n) = 2 i \bigg[ \frac{\partial}{\partial \alpha} \mathcal{I}_{GR}(n) + \frac{\partial^3}{\partial \alpha^3} \mathcal{I}_{GR}(n) \bigg].
\end{equation}

This integral only appears as $\mathcal{I}_2(1)$, which evaluates to the relatively simple expression
\begin{equation}
\mathcal{I}_2(1) = 2 i \frac{\alpha \delta \big( 3 A \cos A + (A^2 - 3) \sin A \big)}{A^5}.
\end{equation}

\subsubsection{$\mathcal{I}_3(n)$}

The remaining integral, $\mathcal{I}_3$, is the simplest to calculate:
\begin{equation}
\mathcal{I}_3(n) = - 2 \frac{\partial^2 \mathcal{I}^*}{\partial \alpha^2}.
\end{equation}

This integral only appears as $\mathcal{I}_3(0)$, which can be evaluated to yield
\begin{equation}
\mathcal{I}_3(0) = \frac{6 \alpha^2 \cos A}{A^4} - \frac{2 \cos A}{A^2} + \frac{6 \delta^2 \sin A}{A^5} - \frac{4 \sin A}{A^3} + \frac{2 \alpha^2 \sin A}{A^3}.
\end{equation}

\clearpage


\clearpage


\end{document}